\documentclass[aps, a4paper, twocolumn, superscriptaddress,10pt, accepted=2021-12-10]{quantumarticle}
\pdfoutput=1

\usepackage{textcomp,hyperref,enumerate, url, cancel}                    			
\usepackage[margin=0.8in]{geometry}
\usepackage[]{xcolor}
\usepackage{amsmath,amssymb,amsthm,bbm,bm,nicefrac}   				
\usepackage{graphicx,epsfig}                                                               		
\usepackage{multirow}

\newcommand{\ket}[1]{\left|#1\right\rangle}							
\newcommand{\bra}[1]{\left\langle#1\right|}
\newcommand{\proj}[1]{\ket{#1}\!\!\bra{#1}}
\newcommand{\ketbra}[2]{\left|#1\rangle\langle#2\right|}

\newcommand{\abs}[1]{\left\lvert #1\right\rvert}

\newcommand{\dketbra}[1]{\left|#1\right\rangle\left\langle#1\right|}

\newcommand{\be}{\begin{equation}} 							
\newcommand{\ee}{\end{equation}}
\newcommand{\ba}{\begin{align}}
\newcommand{\ea}{\end{align}}
\newcommand{\bematrix}{\left(\begin{matrix}}
\newcommand{\ematrix}{\end{matrix}\right)}

\theoremstyle{definition}

\theoremstyle{theorem}
\newtheorem{theorem}{Theorem}[section]

\theoremstyle{lemma}

\theoremstyle{proposition}
\newtheorem{proposition}[theorem]{Proposition}

\theoremstyle{corollary}

\theoremstyle{observation}

\theoremstyle{remark}

\newcommand{\figref}[1]{Fig.~\ref{#1}}

\def\R{{\ensuremath{\mathbbm R}}}

\def\ii{\mathrm{i}}
\newcommand{\tr}[1]{\mathrm{Tr}\left[#1\right]}

\def\d{\mathrm{d}}

\def\cA{\mathcal A}											
\def\cB{\mathcal B}

\def\cE{\mathcal E}

\def\cH{\mathcal H}

\def\cK{\mathcal K}
\def\cL{\mathcal L}
\def\cM{\mathcal M}

\def\cS{\mathcal S}

\newcommand{\eqq}[1]{Eq.~\eqref{#1}}

\usepackage[normalem]{ulem}

\begin{document}
\title{\centering Squeezing-enhanced communication \newline without a phase reference} 
\author{M. Fanizza}\email{marco.fanizza@uab.cat}
\affiliation{NEST, Scuola Normale Superiore and Istituto Nanoscienze-CNR, I-56126 Pisa, Italy}
\affiliation{F\'isica Te\`orica: Informaci\'o i Fen\`omens Qu\`antics, Departament de F\'isica, Universitat Aut\`onoma de 
Barcelona, 08193 Bellaterra (Barcelona) Spain}
\author{ M. Rosati}\email{matteo.rosati@uab.cat}
\affiliation{F\'isica Te\`orica: Informaci\'o i Fen\`omens Qu\`antics, Departament de F\'isica, Universitat Aut\`onoma de 
Barcelona, 08193 Bellaterra (Barcelona) Spain}
\author{M. Skotiniotis}
\affiliation{F\'isica Te\`orica: Informaci\'o i Fen\`omens Qu\`antics, Departament de F\'isica, Universitat Aut\`onoma de 
Barcelona, 08193 Bellaterra (Barcelona) Spain}
\author{J. Calsamiglia}
\affiliation{F\'isica Te\`orica: Informaci\'o i Fen\`omens Qu\`antics, Departament de F\'isica, Universitat Aut\`onoma de 
Barcelona, 08193 Bellaterra (Barcelona) Spain}
\author{V. Giovannetti}
\affiliation{NEST, Scuola Normale Superiore and Istituto Nanoscienze-CNR, I-56126 Pisa, Italy}

\begin{abstract}
We study the problem of transmitting classical information using quantum Gaussian states on a family of phase-noise channels with a finite decoherence time, such that the phase-reference is lost after $m$ consecutive uses of the transmission line. This problem is relevant for long-distance communication in free space and optical fiber, where phase noise is typically considered as a limiting factor. The Holevo capacity of these channels is always attained with photon-number encodings, challenging with current technology.  Hence for coherent-state encodings the optimal rate depends only on the total-energy distribution and we provide upper and lower bounds for all $m$, the latter attainable at low energies with on/off modulation and photodetection. We generalize this lower bound to squeezed-coherent encodings, exhibiting for the first time to our knowledge an unconditional advantage with respect to any coherent encoding for $m=1$ and a considerable advantage with respect to its direct coherent counterpart for $m>1$. This advantage is robust with respect to moderate attenuation, and persists in a regime where Fock encodings with up to two-photon states are also suboptimal. Finally, we show that the use of part of the energy to establish a reference frame is sub-optimal even at large energies. Our results represent a key departure from the case of phase-covariant Gaussian channels and constitute a proof-of-principle of the advantages of using non-classical, squeezed light in a motivated communication setting.
\end{abstract}
\maketitle

\section{Introduction}
The ability to establish and maintain a shared reference frame~\cite{Bartlett2007} between the sender and receiver is often an implicit assumption in communication scenarios.
This is the case, for example, in long-distance communication on optical fiber and in free space, where the information is encoded into quantum states of the electromagnetic field~\cite{Caves1994}. The most fundamental noise models in this scenario are phase-covariant Gaussian channels~\cite{Holevo2012,holevoBOOK}, such as the thermal attenuator and the additive-noise channel~\cite{Holevo2012,holevoBOOK}. For these channels coherent states are known to minimize the output entropy~\cite{Giovannetti2004,Mari2014,Giovannetti2014,Giovannetti2015} and can be used to { design codes attaining {attain}} the channel capacity, i.e., the maximum classical-information transmission rate, provided that the sender and the receiver share a phase reference. Therefore, in practice, typical communication protocols {rely on establishing a common phase reference by sending a high energy coherent state}. Once the common reference is established, information can be encoded into the amplitude and the phase of a coherent state~\cite{Guha11,Rosati16c,Banaszek2020}, easily generated by a classical source. Clearly, then, the use of non-classical sources in this setting provides no communication advantage.

However, {encoding on the phase degree of freedom can be greatly affected if the sender and receiver cannot maintain the common phase reference during the communication protocol {cannot phase-lock their signals}}. This can happen when the relative phase drifts during transmission due to a physical mechanism in the medium, e.g., Kerr non-linearities and temperature fluctuations in optical fiber~\cite{Gordon1990,Wanser1992,Kunz2018} or turbulence effects in free space~\cite{Sinclair2014}; but it can also be an effective result of other mechanisms, e.g., the use of a photodetector to measure the signals or the presence of a malicious eavesdropper~\cite{Diamanti2015,Qi2015}. Several works analyzed the effect of phase noise on common communication methods based on coherent states encodings~\cite{Olivares2013,Jarzyna2014,Jarzyna2016,Adnane2019,DiMario2019,Chesi2018}. 
\\
 \begin{figure}[tb!]
    \includegraphics[trim={0 .2cm 0 0},clip,width=.4\textwidth]{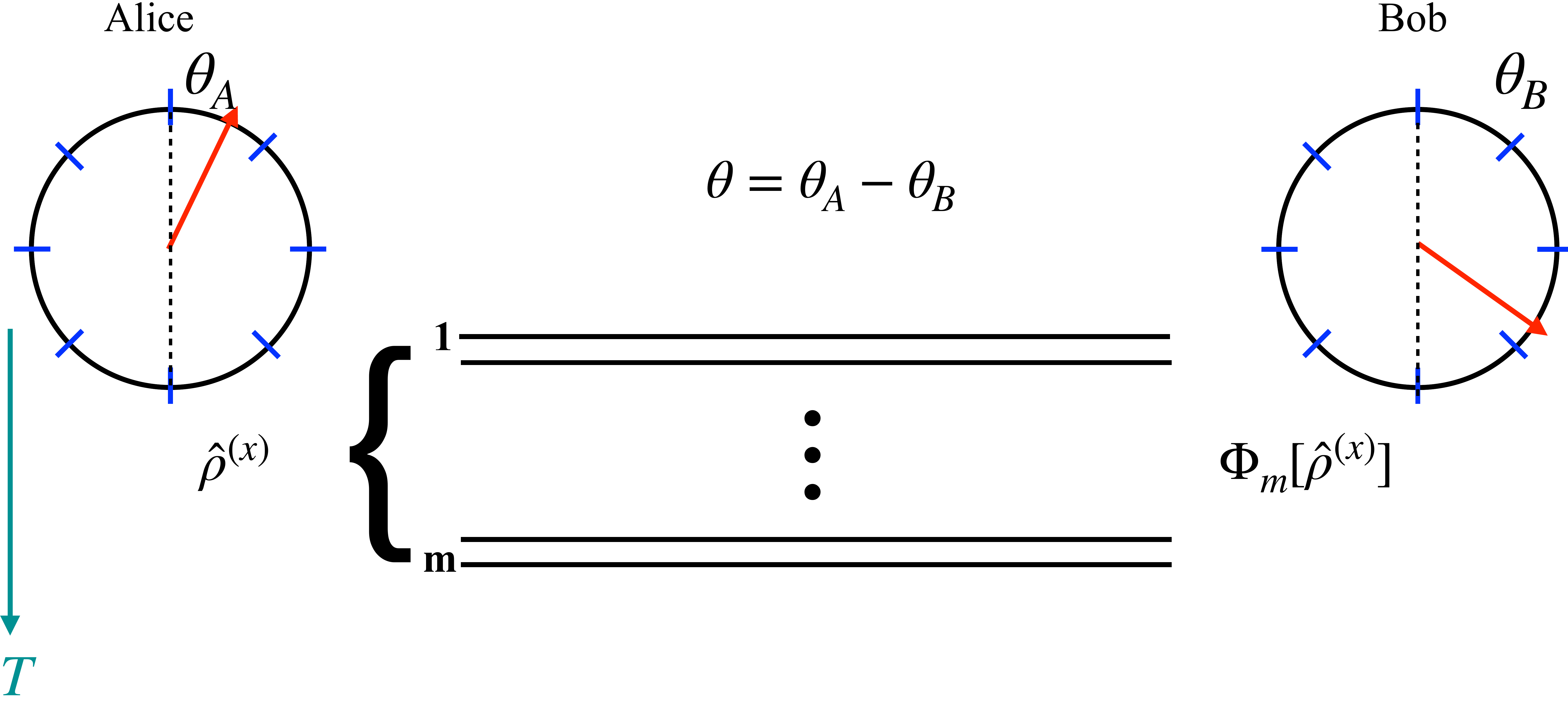} 
    \quad
    \includegraphics[trim={0 .2cm 0 0},clip,width=.4\textwidth]{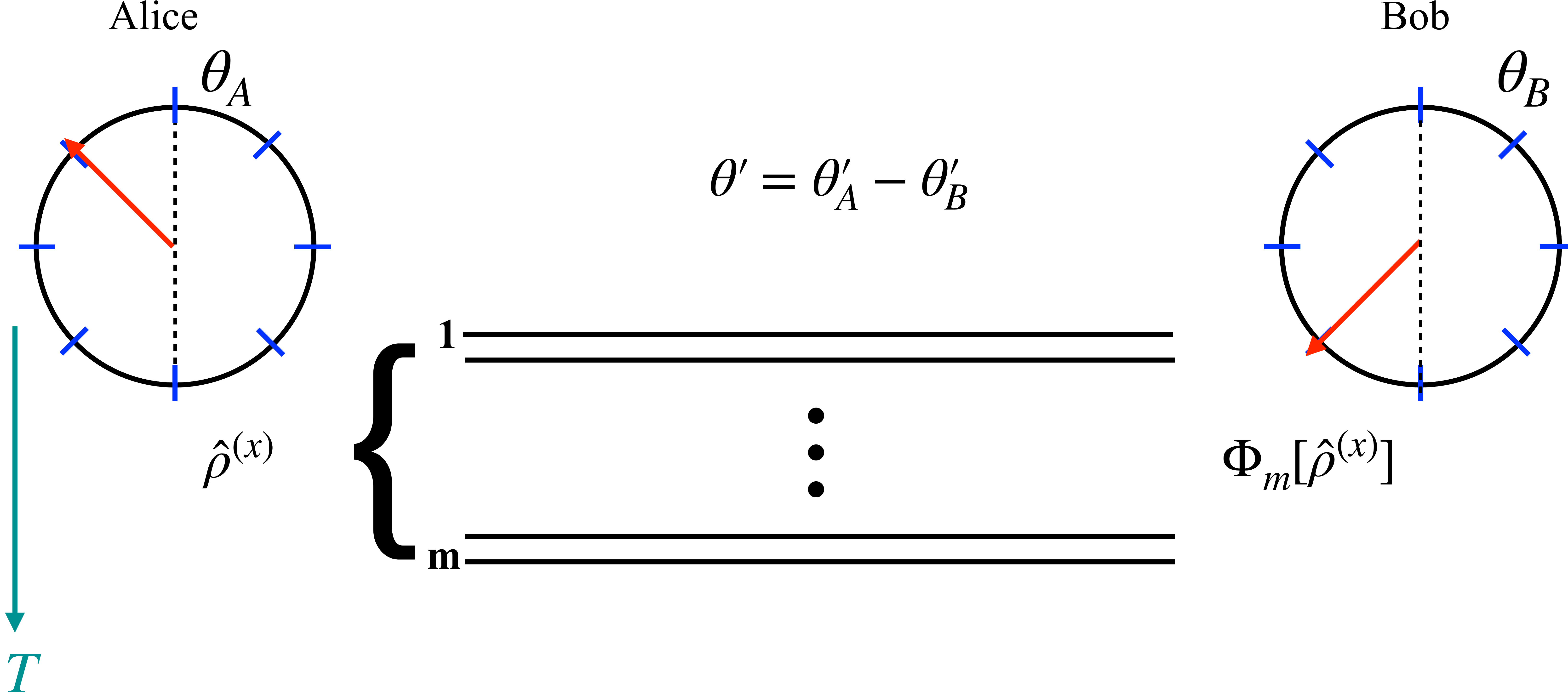}
    \caption{Two uses of the phase-noise memory channel $\Phi_{m}$ analyzed in this work, modeling communication in the absence of a phase-reference.  
    The relative phase $\theta\in(0,2\pi]$ between Alice and Bob is fixed { for a time $T=m\delta t$} but unknown.  Alice exchanges a total of $m$ pulses with Bob, 
    each lasting a time $\delta t$, in the form of a quantum state $\hat\rho^{(x)}\in\cB(\cL^2(\R^{m}))$ encoding the classical message $x\in X$.  After a time 
    $T$, the relative phase between Alice and Bob's local oscillators changes as a result of each party's local oscillator phase-drifting.}
    \label{fig:channel}
\end{figure}
 
In this work we address the question of reliable transmission of classical information for non-Gaussian memory channels~\cite{Caruso2014} that describe the lack of a common frame of reference (see Fig.~\ref{fig:channel}).  Specifically, we focus on the realistic scenario where decoherence takes place in a finite 
time $T$~\cite{Gordon1990,Wanser1992,Sinclair2014,Jarzyna2014,Kunz2018} during which the sender can send a total of 
$m=\lfloor\frac{T}{\delta t}\rfloor$ signals before the onset of decoherence~\footnote{An alternative is to send a finite number of 
signals in parallel using different frequency slots.}, where $\delta t$ is the duration of each signal.  Whilst it is simple to show that the classical 
capacity for such channels can be achieved using Fock states, here we are interested in determining the maximum rate achievable via more experimentally friendly encoding schemes that make use of Gaussian states. Aside for the obvious practical relevance, the problem is interesting 
also theoretically as most studies focus on Gaussian channels.

\begin{figure}[t!]\centering
\includegraphics[trim={ 0cm, 4cm, 0cm, 3cm},clip,width=.45\textwidth]{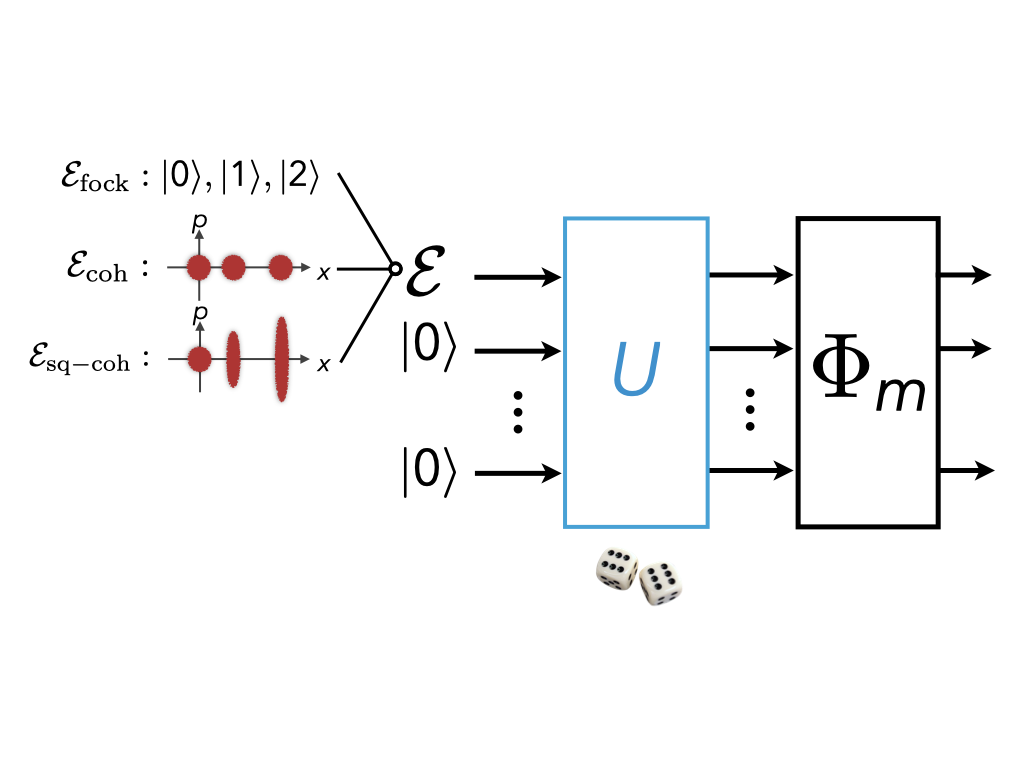} 
\caption{Depiction of several communication strategies on phase-noise memory channels studied in the article. Covariant strategies employ a Haar-random passive interferometer $\hat U$ to increase the communication rate of any initial ensemble. The input ensembles we consider are discrete constellations comprising the vacuum state and one or more pulse signals, including Fock, coherent and squeezed-coherent states.}
\label{fig:strategies}
\end{figure}

Aside from the restriction to Gaussian states, there are additional practical constraints that need to be taken into account when determining the classical 
transmission rate of a quantum channel.  For instance, fibre optical transmission lines have a maximum power cut-off, whereas free space optical links are 
often limited by the power of the source used.  As a result, one seeks to maximize the transmission rate of a channel subject to a suitable constraint that 
depends on the resources at hand.  Throughout this work we will be interested in transmission rates subject to an average energy constraint and explore 
strategies that utilize Gaussian state encodings or Fock state with low photon number obeying an average energy constraint (see Fig.~\ref{fig:strategies} for 
a depiction of some analyzed strategies). 

In absence of loss, the channel we consider is a special case of the channel studied in~\cite {Jarzyna2014}, which considers a more general phase noise. There, the authors derived approximations to achievable rates of coherent-state encodings with fixed energy in the low-photon-number sector. Here instead, in the particular case we consider, we analyze the performance of more general Gaussian encodings, we find exact upper bounds for the optimal coherent state rates, characterize the optimal coherent state rate at low and high energy, evaluate rates of explicit encodings, and we consider losses.

In particular, we show that the addition of squeezing can greatly enhance the communication rate compared to coherent-state encodings, for this family of phase-noise channels.
In particular, in the case of complete dephasing ($m=1$, coinciding with the channel seen by a photodetector), we exhibit an explicit squeezed-coherent-state encoding whose rate surpasses any coherent-state communication strategy for suitable values of the average energy of the signals. Our results provide a clear departure from the case of phase-covariant Gaussian channels and prove the unconditional advantage of using non-classical Gaussian light in a physically motivated communication setting. We note that several works observed an enhancement in discrimination and communication in presence of phase noise via the addition of squeezing to coherent states, though with several restrictions on sources and measurements~\cite{Yuen1978, Saleh1987, Vourdas1994, Yuen2004, Cariolaro2014, Chesi2018}. To our knowledge, our work is the first to prove an unconditional advantage.

Furthermore, we show that the advantage of squeezed-coherent encodings is robust with respect to the addition of channel losses, contrarily to Fock-state encodings, identifying a regime where the use of a source producing up to two-photon Fock states is sub-optimal with respect to a much simpler squeezed-coherent source with reasonable squeezing values, e.g., $\sim$ 5.8dB at energy $E\sim 2$ vs the 8-15dB attainable at the state of the art~\cite{Vahlbruch2016, Zhang2021}.

Finally, we show that the use of part of the signals to establish a common phase reference~\cite{Shapiro1991,Berry2000,Bartlett2007} on these channels is in general detrimental for the communication rate, even at large signal energies.

The rest of the article is structured as follows: in Sec.~\ref{sec:math} we recap the general communication problem and introduce the channel model; in Sec.~\ref{sec:encodings} we first compute the channel capacity, showing it is attained by Fock encodings, then prove the optimality of covariant encodings, 
and finally restrict to Gaussian encodings; in Sec.~\ref{sec:bounds} we first determine upper bounds on coherent-state encodings, applying recent advances in bounding the classical Poisson channel capacity~\cite{Wang2014,Cheraghchi2018}, then we compute attainable lower bounds on coherent and squeezed-coherent encodings via explicit discrete-pulse alphabets; moreover, in Sec.~\ref{sec:loss} we tackle the case of loss and phase-noise, while in Sec.~\ref{sec:phaseRef} we prove the strict sub-optimality of phase-estimation strategies at large energies; finally, in Sec.~\ref{sec:conclusions} we discuss the relevance of our results.

\section{Communication in the absence of a phase reference}\label{sec:math}
In this section we introduce the mathematical model describing the effective channel associated with the lack of a common phase reference, and establish the notation used throughout the remainder of our work.
  
\subsection{Phase-noise memory channels}
Consider two parties---a sender A and a receiver B---who wish to communicate classical information using quantum states of 
light (see \figref{fig:channel}). Both A and B are in possession of their own phase reference in the form of local oscillators, i.e., high-intensity laser light with respective phases $\theta_S$ and $\theta_R$; in general $\theta_S\neq\theta_R$. { Consequently, any state 
$\hat\rho\in\cB(\cL^2(\R^{m}))$, prepared by A is described from B's point of view as $e^{\ii\theta\hat{n}} \hat\rho e^{-\ii\theta\hat{n}}$, where $\theta=\theta_S-\theta_R$, $\hat n=\sum_{i=1}^{m}\hat{n}_i$ is the total-photon-number operator, $\hat{n}_{i}=\hat a_{i}^\dag \hat a_{i}$ where $\hat a_{i},\, \hat a_{i}^{\dag}$ are the bosonic 
creation and annihilation operators satisfying the canonical commutation relations $[\hat a_{i},\hat a_{j}^{\dag}]=\delta_{ij},[\hat a_{i},\hat a_{j}]=0$.  Thus, the phase mismatch 
between A and B's phase references is operationally equivalent to a phase-shift. 

In the simplest case, i.e., $m=1$ in \figref{fig:channel}, the relative phase between A and B, $\theta=\theta_S-\theta_R$, drifts after each use of the transmission line.}  If this  {\it relative phase shift} is fixed but otherwise 
unknown then B's description of the state sent by A is 
\begin{align}
{\Phi_1}(\hat\rho)&:=\int_0^{2\pi}\, \frac{\d\theta}{2\pi}\, e^{\ii\theta\hat{n}_{1}} \hat\rho e^{-\ii\theta\hat{n}_{1}}\nonumber\\&=\sum_{n_{1}=0}^\infty \proj{n_{1}}  \hat\rho \proj{n_{1}},
\label{eq:phasetwirl}
\end{align}
where $\ket{n}$ is a Fock state of the light mode, such that $\hat n_{{1}}\ket n=n\ket n$. That is, the lack of a common phase reference is operationally equivalent to a completely dephasing channel.
We note that the latter describes also the situation where the channel between A and B is ideal, and B decodes all signals sent by A 
using a photodetector.

However, {as mentioned in the introduction {in practice}}, any phase-variation mechanism will take place in a finite amount of time $T$, hence the unknown relative phase, $\theta\in(0,2\pi]$, will remain fixed during this period. Therefore, A and B are capable of exchanging $m=\lfloor\frac{T}{\delta t}\rfloor$ signals, if each of them has duration $\delta t$. We model each signal as a different light mode so that the effective channel we consider takes as input a state, $\hat\rho\in\cB(\cL^2(\R^{m})){=:\cH}$,
and applies the same fixed, but otherwise random, phase $\theta$ on all $m$ modes, i.e.,   
\be
\Phi_{m}(\hat\rho):=\int_0^{2\pi} \frac{\d\theta}{2\pi}\, e^{\ii \theta\hat{n}}\, \hat\rho\, e^{-\ii \theta\hat{n}}=\sum_{n=0}^{\infty}p(n)\hat\rho_{n}
\label{eq:channel}
\ee
where {{$\hat n=\sum_{l=1}^{m}\hat{n}_l$ is the total-photon-number operator,}} $p(n):=\tr{\hat\Pi_n\hat\rho}$, 
$\hat\rho_n:=\hat\Pi_{n}\hat\rho\hat\Pi_n/p(n)$ and $\hat \Pi_{n}$ is the projector on the subspace of $\cH$ with total photon number $n$.  Therefore, the channel $\Phi_{m}$ qualifies as a non-Gaussian memory channel. 

Observe that whereas A and B cannot use the global-phase degree of freedom of light to encode information---there are no coherences between states with different total photon number---they can still utilise the relative phase between
$m$-mode states with a fixed total photon number since $\Phi_{m}$ commutes with the action of energy-preserving Gaussian 
unitaries, i.e., $m$-mode passive interferometers. 

One may wonder why A and B don't simply use some of their resources to perform a standard phase estimation in order to
{ establish a common phase-reference {phase-lock their lasers}}~\cite{Berry2000} and then perform a standard communication protocol. As we will show in Sec.~\ref{sec:phaseRef}, this possibility is sub-optimal in practice when A and B's average energy 
resources are finite. This notwithstanding, {this strategy {phase-locking}} suffers from another obvious shortcoming, as A and B would need to repeat such procedure at regular time intervals $T$.

\subsection{The capacity problem}\label{subsec:capacity}
The classical capacity of a quantum channel is defined as the maximum reliable transmission rate of classical information attainable by employing the channel asymptotically many times with arbitrary encoding and decoding operations~\cite{holevoBOOK}. The capacity of a channel $\Phi$ from Hilbert space $\cH$ to Hilbert space $\cK$ is given by the regularized formula
\begin{equation}
C(\Phi):=\lim_{k\rightarrow\infty}\frac1k\max_{\mathcal{E}_{k}}\chi(\Phi^{\otimes k},\mathcal{E}_{k}),
\label{eq:capacity}
\end{equation}
where the maximization is over all input ensembles $\mathcal{E}_{k}=\{q(x),\hat\rho_{k}^{(x)}\}$, with 
{${\hat\rho_{k}^{(x)}}$ states of the Hilbert space $\cH^{\otimes k}$,} ${{\hat\rho_{k}^{(x)}}\in\cH^{\otimes k}}$ a set of (possibly entangled) states across $k$ independent uses
of the channel, and $x\in X$ is a label which parametrizes the set of signal states on $\mathcal H ^{\otimes k}$. Here 
\begin{align}\nonumber
\chi(\Phi, \cE):&=S\left(\int dx\, q(x) \Phi(\hat\rho^{(x)})\right)\\
&-\int dx\, q(x) S(\Phi(\hat\rho^{(x)})),
\label{eq:Holevo}
\end{align}
is the Holevo quantity~\cite{holevoBOOK}, with $S(\cdot)$ the von Neumann entropy. 
Note that, for infinite-dimensional Hilbert spaces, one further needs to constrain the signals in order to avoid a divergence of the transmission rate. Typically, this is done by enforcing an average-energy constraint on the input ensemble: $\int dx q(x) \hat\rho^{(x)}\leq E$, so that the information transmission rate is a function of $E$. In optical communication, this constraint is motivated by a practical limit in the source power. 

The achievability of \eqref{eq:capacity} requires in general the use of entangled inputs on $\cH^{\otimes k}$~\cite{hastings2009}. However, for phase-covariant Gaussian channels it was recently shown~\cite{Mari2014,Giovannetti2014,Giovannetti2015} that the regularization is not necessary and the capacity is given by the maximization of the single-letter formula \eqref{eq:Holevo}, attainable with coherent-state signals. As we will show in Sec.~\ref{subsec:fock}, this is the case also for our phase-noise memory channel $\Phi_{m}$ by using Fock signals.

Importantly, entanglement can be used at the output of the channel as well. Indeed, if one is restricted to use a specific type of signals $\hat\rho^{(x)}\in\cS$, then 
\begin{equation}\label{eq:constrainedRate}
R_{\cS}(\Phi):=\max_{\cE\text{ s.t. }\forall x\, \hat\rho^{(x)}\in\cS}\chi(\Phi,\cE)
\end{equation} 
determines the maximum rate attainable by sending sequences of signals extracted from $\cS$ over multiple uses of the channel and decoding them with an optimal collective quantum measurement~\cite{holevoBOOK}. The realization of such measurement is still not trivial in practice, even for coherent-state codes~\cite{Guha11,Rosati16b,Rosati16c,Rosati2017,Banaszek2020}, and typical low-energy communication methods commit to specific single-system quantum measurements $\cM:=\{ \hat M_{y}\geq0,\, \sum_{y} \hat M_{y}=I\}$, which, in conjunction with a specific type of signals $\cS$, induce a classical channel with maximum information transmission rate given by its Shannon capacity:
\begin{equation}\label{eq:shannonRate}
\small R_{\cS}(\Phi;\cM):=\max_{{\bf q}}\sum_{x,y}q(x)p^{(x)}(y)\log\frac{p^{(x)}(y)}{\sum_{x'}q(x') p^{(x')}(y)},
\end{equation}
where the term to maximize is the classical mutual information between the input $x$, with probability distribution ${\bf q}$, and the output $y$, with conditional probability distribution ${\bf p}^{(x)}$ such that ${\bf p}^{(x)}(y):=\tr{\hat M_{y}\Phi(\hat\rho^{(x)})}$.

\section{Phase-noise channel: capacity, covariant and Gaussian rates}\label{sec:encodings}
In this section we compute the phase-noise memory channel capacity and several maximum information transmission rates with Gaussian encodings. We make use of covariant encodings~\cite{Korzekwa2019}, which randomize any given encoding, and we show that they attain larger or equal rate for our channel.

\subsection{Classical capacity of the phase-noise channel and Fock encodings}\label{subsec:fock}

In this section we show that the classical capacity of {the effective channel} $\Phi_m$ is $C(\Phi_{m},E)=m\,g(\frac Em)$, where $g(E):=(E+1)\log(E+1)-E\log E$ is the entropy of a single-mode thermal state of average energy $E$. 

It is straightforward to see that for an ensemble $\mathcal{E}_{k}$ with energy constraint $kE$ 
\be\label{UBCap}
\chi(\Phi_{m}^{\otimes k},\mathcal{E}_k)\leq S(\hat\rho_{\rm th}(kE))=\sum_{j=1}^{mk}g(E_{j})= k\,m\,g\left(\frac Em\right),
\ee
where the first inequality follows from discarding negative terms in the Holevo quantity and using the fact that the entropy is maximized, under an average-energy constraint, by a thermal state~\cite{Yuen1993,Caves1994}. This follows from $0\leq D(\hat\rho||\hat\rho_{\rm th}(E))=S(\hat\rho_{\rm th}(E))-S(\hat\rho)$ whenever $\tr{\hat\rho\hat H}=E$, with equality if and only if $\hat\rho=\hat\rho_{\rm th}(E)$. Here $H$ is the total photon number of $mk$ modes and $\hat\rho_{\rm th}(kE)$ is a thermal state of $mk$ modes with total average energy $kE$, which can always be written as tensor product of single-mode thermal states with average energies $E_j=E/m$. Finally, $k\,m\,g(\frac Em)$ is monotonic in $E$, therefore it is not restrictive to constrain the total energy to be $kE$.

Now note that the upper bound of \eqq{UBCap} is achievable using an ensemble of tensor-product Fock states on $m$ modes, i.e., mapping $x\in X\mapsto \bigotimes_{i=1}^{m}|n_{i}^{(x)}\rangle$. Indeed, Fock states are pure, giving a zero contribution to the second term of the Holevo quantity, and invariant under the action of $\Phi_{m}$, so that with a thermal probability distribution of total average energy $E$, the average output state of the channel is exactly $\hat\rho_{\rm th}(E)$. Hence we conclude that $C(\Phi_{m},E)=m\,g(\frac Em)$. Moreover, the same arguments above apply to any phase-noise channel with arbitrary phase distribution, {{provided that the phase-shift is identical on each mode,}} so that their capacity is given by the same expression.

We stress that this is the same rate attainable by $m$ uses of the identity channel with average energy per mode $\frac Em$, implying that, if the sender and receiver can produce and detect Fock states, then $\Phi_{m}$ is essentially noiseless. \\

Finally, as Fock states with increasing photon number are increasingly difficult to produce, it makes sense to 
consider the rate obtainable by sending ensembles of Fock states with fixed maximum photon number. Repeating the proof of this section, the optimal rate of these ensembles as defined in \eqref{eq:constrainedRate} is readily characterized as 
\begin{equation}
R_t(E)=\max_{s<E} g(s,t),
\end{equation} 
where $g(s,t)=\beta(s)s -\log\left(\frac{1-e^{-\beta(s)(t+1)}}{1-e^{-\beta(s)}}\right)$ is the entropy of the thermal state of the truncated Hilbert space with photon number up to $t$, with average photon number $s$, with inverse temperature $\beta(s)$.

Note that for any ensembles of $t$ states the maximum rate is bounded by the maximum value of the Holevo quantity, i.e., $\log t$. For restricted Fock ensembles with photon number up to $1$ or $2$ states the Holevo quantity saturates to respectively $\log 2$ and $\log 3$ as the energy constraint grows.

\subsection{Covariant encodings}
Since photon-number states are hard to produce, one can be interested in constraining the ensembles to more accessible states~\footnote{Note that, although the channel is additive, i.e., its capacity is attained with product encodings over different uses, superadditivity may arise due to the constrained input. For simplicity, we will restrict to product encodings in the paper.}. A drastic simplification in the optimization over any family of states can be obtained by exploiting the symmetry of the channel {(see \cite{Korzekwa2019} for a general resource-theoretic treatment of encoding-restricted communication)}. 

We use the fact that the average on Haar-random energy-preserving Gaussian unitaries $\hat{U}$ (or passive interferometers PI), which act as the group ${\rm U}(m)$ in phase space~\cite{Serafini2007}, completely depolarizes the state in blocks of fixed total photon number $n$, which have dimension $\binom{n+m-1}{m-1}$: 
\begin{equation}\label{eq:twirling}
\int_{{\rm U}(m)}dU \, \hat U \hat\rho \hat U^\dagger = \sum_{n=0}^{\infty} p(n)  \frac{\hat{\Pi}_n}{\binom{n+m-1}{m-1}}.
\end{equation}
This is a consequence of Schur's lemma applied to the decomposition into irreducible representations of ${\rm U}(m)$ of the Hilbert space of $m$ modes. The decomposition in turn can be understood as a consequence of the connection between coherent states of an infinite-dimensional system with spin-coherent states of finite dimension~\cite{Perelomov1972,Zhang1990}, detailed in the Appendix~\ref{coherentgroups}. 

Exploiting this property, the rate achievable with an arbitrary ensemble $\cE=\{q(x),\hat\rho^{(x)}\}$ is then bounded by
\be\begin{aligned}\label{eq:rateDef}
&\chi(\Phi_{m},\cE)\leq \Bigg[S\left(\int dx\, q(x)  \int_{\mathrm U(m)} dU \Phi_m(\hat U\hat\rho^{(x)} \hat U^\dagger) \right)\\
&-\int dx\, q(x)  \int_{\mathrm U(m)} dU S(\Phi_m(\hat U\hat\rho^{(x)} \hat U^\dagger))\Bigg]\\
&=\chi(\Phi_{m},\cE^{\rm Haar})
\end{aligned}\ee
where the inequality follows from the concavity and unitary-invariance of the von Neumann entropy and the fact that $\hat U$ and $\Phi_m$ commute, while in the last equality we defined $\cE^{\rm Haar}$ as the ensemble obtained by applying a Haar-random PI $\hat U$ to the states extracted from $\cE$.  The inequality means that one can always restrict the maximization of the Holevo quantity to ensembles of the form $\cE^{\rm Haar}$, which are invariant under total-phase shifts and thus constitute what we refer to as \emph{covariant encodings}.

It follows that for any ensemble of states $\cE$ with total photon number distribution 
${\bf p}^{(x)}$, where ${\bf p}^{(x)}(n)=\tr{\hat\Pi_n \hat\rho^{(x)}}$ {and writing  $\hat\rho^{(x)}_n:=\hat\Pi_{n}\hat\rho^{(x)}\hat\Pi_n/\tr{\hat\Pi_{n}\hat\rho^{(x)}}$,} use of \eqref{eq:twirling} reduces the Holevo quantity to 
\be\begin{aligned}\label{eq:maxRateGen}
&\chi(\Phi_{m},\cE^{\rm Haar})=\Bigg[H\left(\int dx\, q(x) {\bf p}^{(x)}\right)\\
&-\int dx\, q(x) H({\bf p}^{(x)})+\sum_{n=0}^{\infty}\int dx\, q(x) {\bf p}^{(x)}(n)\\
&\times\left(\log\binom{n+m-1}{m-1}-S(\hat \rho_n^{(x)})\right)\Bigg]\\
&=mg\left(\frac{E}{m}\right)-D\left(\int dx\, q(x) {\bf p}^{(x)} || {\bf p}^{\rm th}\right)\\
&-\int dx\, q(x) \left(H({\bf p}^{(x)})+\sum_{n=0}^{\infty}{\bf p}^{(x)}(n)S(\hat \rho_n^{(x)})\right),
\end{aligned}\ee
where $D(\cdot || \cdot)$ is the Kullback-Leibler divergence, $H(\cdot)$ the Shannon entropy, and 
\be
{\bf p}^{\rm th}(n)=\binom{n+m-1}{m-1}\left(\frac{E}{E+m}\right)^n \left(\frac{m}{E+m}\right)^m
\label{eq:thermal}
\ee
is the total-photon-number distribution of the thermal state on $m$ modes with average energy per mode $\frac Em$. 

From this expression it is intuitively apparent that states with peaked total-photon-number distribution are preferable as they make $H({\bf p}^{(x)})$ 
smaller without necessarily increasing $D(\int dx\, q(x) {\bf p}^{(x)} || {\bf p}^{\rm th})$. Indeed, as already mentioned above, the capacity 
of the channel is attained by a thermal ensemble of Fock states. This fact will be crucial in understanding why squeezed states offer 
an enhancement (see Sec.~\ref{subsec:squeezing}). 

\subsection{Gaussian encodings}
In the rest of the article we will restrict to Gaussian encodings, which are easily realizable in practice. Note that, thanks to the concavity of the entropy and the fact that any Gaussian state can be written as a mixture of pure Gaussian states, it is straightforward to further restrict the optimization to pure Gaussian states (see Appendix~\ref{puregauss}). 
The position and momentum operators of $m$ bosonic modes can be grouped as a vector $\mathbf{\hat{r}}=(\hat{x}_1,\hat{p}_1,...,\hat{x}_m,\hat{p}_m)$. These operators satisfy the canonical commutation relations $[\hat{r}_i,\hat{r}_j]=i\Omega_{ij} I$, where 
\be
\Omega:=\bigoplus_{i=1}^{m}\left(\begin{matrix}
0&1\\-1&0\end{matrix}\right). 
\label{eq:symplecticform}
\ee
A key class of Gaussian unitary operators is that of displacements
\be
\hat D({\mathbf s})=\exp\left(-\ii {\bf s}^{\sf T} \Omega \hat{\bf r}\right), \qquad \mathbf{s}\in {\mathbb R}^{2m}, 
\label{eq:dispacement}
\ee
satisfying  $\hat D(\mathbf{s_1+s_2})=\hat D(\mathbf{s_1})\hat D(\mathbf{s_2})e^{\ii \mathbf{s}_1^{\sf{T}}\Omega \mathbf{s}_2/2}$. { The displacement operators can be written in terms of the creation and destruction operators using $\hat a_{i}=(\hat x_{i}+\ii \hat p_{i})/\sqrt2$.} The set of displacements constitutes an orthogonal complete operator set and allows the definition of the characteristic function of a trace class operator $\hat\rho\in\cB(\cL^2(\R^{m}))$ as $\phi_{\hat\rho}(\mathbf s):=\tr{\hat\rho \hat D({-\mathbf{s}})}$~\cite{serafiniBOOK}. Gaussian states are those states whose characteristic function is Gaussian, i.e.
\be
\phi_{\hat\rho}(\mathbf s)=\exp\left(-\frac{1}{4} {\mathbf s}^{\sf T}\Omega^{\sf T} V\Omega {\mathbf s} +\ii {\mathbf s}^{\sf T}\Omega 
\boldsymbol{\mu}\right),
\label{eq:gaussianchar}
\ee 
with $\boldsymbol{\mu}:=\tr{\hat{\mathbf{r}}\hat\rho}$ and $V:=\tr{\lbrace(\hat{\mathbf r}-{\boldsymbol\mu}),(\hat{\mathbf r}-\boldsymbol \mu)^\text{T}
\rbrace\hat\rho}$ the associated statistical mean vector and covariance matrix respectively. Any pure Gaussian state $\ket{\psi}$ can be written as 
\be
\ket{\psi}=\hat U\,\hat S(\mathbf r)\, \hat D(\mathbf s)\ket{0}^{\otimes m} 
\label{eq:Gaussians}
\ee
where $\hat D(\mathbf s)$ is a product of single-mode displacements defined above, while $\hat S(\mathbf r)$ is a product of single-mode squeezing operators, defined as
\begin{align}
\hat S(r_{i})&=\exp\left(\frac {r_{i}}{2}(\hat a_{i}^{2}-\hat a_{i}^{\dag 2})\right),
\label{eq:operations}
\end{align}
{{which can be written in terms of the position and momentum operators using $\hat a_{i}=(\hat x_{i}+\ii \hat p_{i})/\sqrt2$ and its hermitian conjugate $\hat a_{i}^{\dag}$}}, and $\hat U$ an energy-preserving Gaussian unitary operation acting on all $m$ modes.

We are then left to consider Gaussian ensembles of the form $\cE^{\rm Haar}_{\rm G}:=\{q(\mathbf r,\mathbf s)\,dU,\hat U\ket{\mathbf r,\mathbf s}\}$, which can be generated by producing a tensor product of $m$ squeezed-coherent states $\ket{\mathbf r,\mathbf s}:=\hat S(\mathbf r) \hat D(\mathbf s)\ket{0}^{\otimes m}$ with probability $q(\mathbf r,\mathbf s)$ and then acting with a $m$-mode Haar-random PI $\hat U$. Consequently, the optimal Gaussian rate, as defined by Eq.~\eqref{eq:constrainedRate} for $\cS$ being the set of Gaussian states, is obtained by maximizing Eq.~\eqref{eq:maxRateGen} over $q(\mathbf r,\mathbf s)$:
\begin{equation}\begin{aligned}
&R_{\rm G}(\Phi_{m})=\max_{q(\mathbf r,\mathbf s)}\Bigg[H\left(\int d\mathbf r\,d\mathbf s\, q(\mathbf r,\mathbf s) {\bf Q}^{(\mathbf r,\mathbf s)}\right)\\
&-\int d\mathbf r\,d\mathbf s\, q(\mathbf r,\mathbf s) H({\bf Q}^{(\mathbf r,\mathbf s)})\\
&+\sum_{n=0}^{\infty}\int d\mathbf r\,d\mathbf s\, q(\mathbf r,\mathbf s) {\bf Q}^{(\mathbf r,\mathbf s)}(n)\log\binom{n+m-1}{m-1}\Bigg],
\end{aligned}\end{equation}
where ${\bf Q}^{(\mathbf r,\mathbf s)}$ is the total-photon-number distribution of $\ket{\mathbf r,\mathbf s}$ obtainable from that of a single-mode squeezed-coherent state~\cite{Yuen1976,Gong1990} (see Appendix~\ref{app:sqCoh}).

\section{Bounds on Gaussian communication rates}\label{sec:bounds}
In this section we determine upper and lower bounds on Gaussian communication rates on $\Phi_{m}$. 
 We recall that a more general phase-noise model encompassing our $\Phi_{m}$ has been
  studied in~\cite {Jarzyna2014}, where the authors derived approximations to
  the Holevo information of coherent-state ensembles with fixed energy in the
  low-photon-number sector. By restricting our attention to $\Phi_{m}$, we perform a wider analysis, extending the attention to general encodings. In particular, in this section we
  \begin{itemize}
  \item evaluate an exact upper bound for the rate of coherent state encodings;
  \item present explicit strategies with covariant encodings and discrete energy pulses;
  \item study the high and low energy behaviour of the optimal coherent state rate and show how to achieve them;
  \item show the advantage of using squeezing with respect to coherent states.
  \end{itemize}
 
\subsection{Maximum coherent-state rate and its upper bounds}
{If we restrict to coherent-state encodings ($\mathbf{r}=0$ in Eq.~(\ref{eq:Gaussians})), we can provide a general expression for the optimal rate using the fact that the total-photon-number distribution of a coherent state $\ket{\mathbf s}:=\hat D(\mathbf s)\ket{0}^{\otimes m}$ is a Poissonian ${\bf P}^{(s)}$ with probabilities ${\bf P}^{(s)}(n):=e^{-s}\frac{s^{n}}{n!}$ and depends only on its total energy $s:=|\mathbf s|^2$. 

The optimization in Eq.~\eqref{eq:maxRateGen} can be restricted to input distributions on the total energy only, $q(s)$, and the optimal coherent-state rate
\be\begin{aligned}\label{eq:maxRate}
&R_{\rm c}(\Phi_{m},E):=\max_{q(s)}\chi(\Phi_{m},\cE_{\rm c}^{\rm Haar})\\
&=\max_{q(s)}\Bigg[\sum_{n=0}^{\infty}\int_0^{\infty} ds\,q(s){\bf P}^{(s)}(n)\log\binom{n+m-1}{m-1}\\
&+H\left(\int_0^{\infty} ds\, q(s){\bf P}^{(s)}\right)-\int_0^{\infty} ds\,q(s)H({\bf P}^{(s)})\Bigg]
\end{aligned}\ee
is attained by producing a single-mode coherent state of energy $s$ with probability $q(s)$ and distributing it with a Haar-random PI. For $m=1$, only the last two terms of \eqq{eq:maxRate} contribute and we recover the well-known discrete-time classical Poisson channel with input distribution $q(s)$. Its capacity is still an open problem in classical information theory for which only upper and lower bounds are known~\cite{Martinez2007,Lapidoth2008a,Lapidoth2009,Wang2014,Cheraghchi2018}. For $m>1$, the first term adds a positive contribution depending on the number of modes $m$ and the output photon-number distribution. 
 
Employing the connection with the Poisson channel, we can upper-bound the optimal coherent-state rate by bounding separately the expressions in the second and third rows of Eq.~(\ref{eq:maxRate}). 
\begin{proposition}\label{prop:upperbound} For any upper bound $R_{\rm c}(\Phi_{1},E)<f(E)$ (where $R_{\rm c}(\Phi_{1},E)$ is the capacity of the Poisson channel) we have
\begin{equation}\label{eq:uBound}
R_{\rm c}(\Phi_{m},E)<f(E)+ \sum_{n=0}^{\infty} {\bf P}^{(E)}(n) \log\binom{n+m-1}{m-1}.
\end{equation}
\end{proposition}

\begin{proof}
In order to prove Proposition \ref{prop:upperbound} we need to upper bound
 the first term in the coherent-state rate. We observe that 
$\sum_{n=0}^{\infty}{\bf P}^{(s)}(n) \log\binom{n+m-1}{m-1}$ is a concave function of $s$. Indeed its second derivative evaluates to

\be\begin{aligned}
&\frac{d^2}{d^2s}\{\sum_{n=0}^{\infty}{\bf P}^{(s)}(n) \log\binom{n+m-1}{m-1}\}\\&=s^{-2}\sum_{n=0}^{\infty}{\bf P}^{(s)}(n)((s-n)^2-n)\log \binom{n+m-1}{m-1} \\&+s^{-2}\sum_{n=0}^{\infty}{\bf P}^{(s)}(n)( s^2 \log\binom{n+m-1}{m-1}\\& +s (n+1) \log\binom{n+m}{m-1} -s(2s+1)\log\binom{n+m}{m-1})\\
&=\sum_{n=0}^{\infty}{\bf P}^{(s)}(n)(\log\binom{n+m-1}{m-1}\\& +\log\binom{n+m+1}{m-1} -2\log\binom{n+m}{m-1})\\
&\leq\sum_{n=0}^{\infty}{\bf P}^{(s)}(n) \sum_{i=1}^m \log\frac{(n+i)(n+i+2)}{(n+i+1)^2}<0
\end{aligned}\ee
where we used ${\bf P}^{(s)}(n) n^\alpha = s {\bf P}^{(s)}(n-1)  n^{\alpha-1}$, $\log\binom{x+m-1}{m-1}=\sum_{i=1}^{m}\log {\frac{x+i}{i}}$, and $\log \frac{x(x+2)}{(x+1)^2}<0$ by monotonicity of the function $\frac{x}{x+1}$. Therefore, by applying again Jensen's inequality on the integral in $s$, as well as recalling that 
$\int_0^{\infty} ds \,q(s) s=E$, we obtain the following inequality:
\begin{align}
\int_0^{\infty}ds \,q(s)&\sum_{n=0}^{\infty} {\bf P}^{(s)}(n)\log\binom{n+m-1}{m-1}\nonumber\\&\leq \sum_{n=0}^{\infty} {\bf P}^{(E)}(n) \log\binom{n+m-1}{m-1},
\end{align}
Hence the optimal coherent-state rate can be upper bounded for all $E$ and $m$ by Eq.~\eqref{eq:uBound}.\end{proof}

Note that the second term in Eq.~(\ref{eq:maxRate}) equals the Holevo quantity of an ensemble of coherent states at fixed expected value of the energy, which~\cite{Jarzyna2014} evaluated for encodings which are not covariant, therefore suboptimal.
In the rest of the article, we employ two upper bounds based on the bounds of Ref.~\cite{Wang2014} and Ref.~\cite{Cheraghchi2018}, on the capacity of the Poisson channel.} 
Explicitly, the bound $R^{\rm up}_{\rm c}(\Phi_m,E)$ is given by Eq.~\eqref{eq:uBound} with~\cite{Cheraghchi2018}:
\begin{align}\label{eq:ubPoi}
&f(E):=E \log \left(\frac{1+\left(1+e^{1+\gamma}\right) E+2 E^{2}}{e^{1+\gamma} E+2 E^{2}}\right)\\
&+\log \left(1+\frac{1}{\sqrt{2 e}}\left(\sqrt{\frac{1+\left(1+e^{1+\gamma}\right) E+2 E^{2}}{1+E}}-1\right)\right)\nonumber\\
&\geq\max_{q(s)}\left[H\left(\int_0^{\infty} ds\, q(s) {\bf P}^{(s)}\right)-\int_0^{\infty} ds\, q(s) H({\bf P}^{(s)})\right],\nonumber
\end{align}
where $\gamma$ is the Euler-Mascheroni constant.

The bound obtained using~\cite{Wang2014} reproduces correctly the expansion of the {channel capacity $C({\Phi_m,E})$} at the first two leading orders, but it appears to be larger than $R^{\rm up}_{\rm c}(\Phi_m,E)$ everywhere but for extremely low energies. Therefore, we employ the former bound only to derive the low-energy expansion of the coherent-state rate, see Sec.~\ref{lowbound}, while the latter bound is employed throughout the rest of the manuscript.

\subsection{Lower bounds on Gaussian rates via discrete-pulse encodings}\label{subsec:squeezing}

\subsubsection{Randomized on/off modulation (ROOM)}

For general $m$, we can determine an achievable lower bound on both the coherent and squeezed-coherent maximum rates by employing a simple randomized on/off modulation (ROOM) at the encoding: with some probability $p$ we send a Haar-random pulse $\hat U\ket{\mathbf r,\mathbf s}$ and the vacuum otherwise. The resulting lower bound for general $\mathbf r,\mathbf s, p$ respecting the mean-energy constraint is
\begin{align}\label{eq:lowBGen}
&R(\Phi_{m},E;\mathbf r,\mathbf s,p)=p\sum_{n=1}^{\infty}{\bf Q}^{(\mathbf r,\mathbf s)}(n)\log\binom{n+m-1}{m-1}\nonumber\\
&+h\left(1-p+p{\bf Q}^{(\mathbf r,\mathbf s)}(0)\right)\nonumber\\
&+\sum_{n=1}^{\infty}h\left(p{\bf Q}^{(\mathbf r,\mathbf s)}(n)\right)-p\sum_{n=0}^{\infty}h\left({\bf Q}^{(\mathbf r,\mathbf s)}(n)\right)\nonumber\\
&= p\sum_{n=1}^{\infty}{\bf Q}^{(\mathbf r,\mathbf s)}(n)\log\binom{n+m-1}{m-1}\\
&+h\left(1-p+p{\bf Q}^{(\mathbf r,\mathbf s)}(0)\right)+\left(1-{\bf Q}^{(\mathbf r,\mathbf s)}(0)\right)h(p)\nonumber\\
&-p\,h\left({\bf Q}^{(\mathbf r,\mathbf s)}(0)\right),\nonumber
\end{align}
where ${\bf Q}^{(\mathbf r,\mathbf s)}$ is the total-photon-number distribution of a tensor product of squeezed-coherent states $\ket{\mathbf r,\mathbf s}$~\cite{Yuen1976,Gong1990} {and $h(p):=-p\log p$}. In particular, as explained above, for the coherent encoding it always suffices to start with a single-mode pulse, i.e., $\mathbf{r}=0$ and $\mathbf s=(s,0,\cdots,0)$, so that ${\bf Q}^{(\mathbf r,\mathbf s)}$ reduces to a Poissonian ${\bf P}^{(\abs{\alpha}^{2})}$.  

Consequently, in accordance with Eq.~\eqref{eq:constrainedRate}, we define the best lower bounds on the maximum rate of coherent and squeezed-coherent encodings as 
\begin{align}
R^{\rm room}_{\rm c}(\Phi_{m},E)&:=\max_{\mathbf s,p}R(\Phi_{m},E;\mathbf s,0,p),\label{eq:roomCoh}\\
R^{\rm room}_{\rm sc}(\Phi_{m},E)&:=\max_{\mathbf r,\mathbf s,p}R(\Phi_{m},E;\mathbf r,\mathbf s,p),\label{eq:roomSqCoh}
\end{align}
where the optimization is over values of the parameters respecting the energy constraint. Clearly, in a ROOM communication strategy, the larger the energy of the Haar-random pulses is, the smaller is their joint probability $1-p$. This fact is particularly clear for coherent encodings, where, as already noted, the problem depends exclusively on the absolute value of $\mathbf s$. Since the energy constraint $E=(1-p)|\mathbf s|^{2}$ imposes $\mathbf s=\mathbf s_{0}:=(\sqrt{E/p},0,\cdots,0)$, the optimization of Eq.~\eqref{eq:roomCoh} essentially reduces to a single-parameter one:
\be
R^{\rm room}_{\rm c}(\Phi_{m},E):=\max_{\mathbf p\in[0,1]}R(\Phi_{m},E;\sqrt{E/p},0,p).
\ee
On the contrary, for the squeezed-coherent encoding it is difficult to optimize \eqq{eq:roomSqCoh} in general. Numerical evidence suggests to concentrate all the energy in one pulse before $\hat U$, i.e., $\mathbf r=(r,0,\cdots,0)$ and $\mathbf s=(s,0,\cdots,0)$, with $s\in\mathbb R$ and $r>0$. Intuitively, this choice is aimed at reducing the photon-number variance as pointed out after Eq.~\eqref{eq:maxRateGen}. 

\subsubsection{On/off modulation plus photodetection (OOP)}
As explained in Sec.~\ref{subsec:capacity}, attaining the maximum rate $R^{\rm room}_{\rm c/sc}(\Phi_{m},E)$ of a ROOM encoding still requires the realization of collective quantum measurements across several uses of the channel. Hence we consider further a fully explicit communication scheme, employing a non-randomized on/off modulation with coherent, $\cA_{\rm c}:=\{\ket0,\ket{\mathbf s}\}$, or squeezed-coherent signals, $\cA_{\rm sc}:=\{\ket0,\ket{\mathbf r,\mathbf s}\}$, plus on/off photodetection measurements (OOP), $\cM_{\rm pd}:=\{\proj 0,I-\proj 0\}$, whose rate $R^{\rm oop}_{\rm c/sc}(\Phi_{m},E;\cA_{\rm c/sc},\cM_{\rm pd})$ can be computed via Eq.~\eqref{eq:shannonRate}.

\subsubsection{Two or more pulses}
It is clear that the ROOM encoding can be generalized by considering more than one covariant pulse, obtaining similar expressions to Eq.~\eqref{eq:lowBGen}. Providing analytical intuition about the behaviour of these strategies is challenging and beyond the scope of this paper. However, in our numerics we do consider ternary coherent and squeezed-coherent encodings, composed of the vacuum state and two randomized pulses with distinct parameters (see Fig.~\ref{fig:strategies} for a depiction of these strategies { and the Mathematica~\cite{Mathematica} notebook~\footnote{Supplementary files.} for the numerics}). As for ROOM, the maximum rate of these encodings can be obtained by optimizing all the parameters subject to the average-energy constraint.

\subsection{Limiting behaviour of the maximum coherent-state rate}

{\subsubsection{High-energy regime}
Let us discuss the high energy regime for coherent-state strategies. The second term of the upper bound Eq.~(\ref{eq:uBound}) can be further bounded using Jensen's inequality, obtaining the coarser bound
\begin{align}\nonumber
\sum_{n=0}^{\infty} {\bf{P}}^{(E)}(n) \log\binom{n+m-1}{m-1}&\leq \log \binom{E+m-1}{m-1}\\
=(m-1)\log E+ O(1),
\label{eq:coarserUB}
\end{align}
which, together with the well-known fact that at high energies the capacity of the Poisson channel is $\frac{1}{2}\log E+O(1)$~\cite{Cheraghchi2018}, establishes the following: 

\begin{proposition}
The maximum rate of transmission of classical information through the channel $\Phi_m$ using high-energy coherent state encodings is given by 
\begin{equation}\label{highen}
R_{\rm c}(\Phi_{m},E)=\left(m-\frac{1}{2}\right)\log E+ O(1).
\end{equation}

\end{proposition}
We check that this asymptotic rate is achievable with a thermal ensemble of coherent states. An ensemble for which the average state is the thermal state can be obtained by encoding with probability distribution given by a Gamma distribution $q(s)=\left(\frac{m}{E}\right)^m \frac{e^{-\frac{s}{E/m}}s^{m-1} }{(m-1)!}$, indeed
\begin{align}
&\int_0^{\infty} ds \,q(s) {\bf P}^{(s)}(n)\nonumber\\&=\binom{n+m-1}{m-1}\left(\frac{E}{E+m}\right)^n \left(\frac{m}{E+m}\right)^m\nonumber \\&={\bf p}^{\rm th}(n).
\end{align}

For this distribution one needs to evaluate only the average-output-entropy term. From the well-known fact that $H({\bf P}^{(s)})\leq \frac{1}{2}\log 2\pi e(s+\frac{1}{12})$ \cite{Lapidoth2009,Adell2010}, and from Jensen inequality, we have

\begin{align}
\int_{0}^{\infty}ds &\,q(s)H({\bf P}^{(s)})\leq \int_{0}^{\infty}ds \,q(s) \frac{1}{2}\log2\pi e (s+\frac{1}{12})\nonumber\\&\leq \frac{1}{2}\log 2\pi e(E+\frac{1}{12}).
\end{align}
We then obtain a rate 
\begin{align}
R^{\rm th}&=mg(E/m)-\int_{0}^{\infty}ds\, q(s)H({\bf P}^{(s)})\nonumber\\&\geq mg(E/m)-\frac{1}{2}\log 2\pi e (E+\frac{1}{12})\nonumber\\&=\left(m-\frac 1 2\right)\log E+ O(1).
\end{align}

Moreover, by fixing $E/m=k$ and sending $m$ to infinity, we get a rate per use of the transmission line which approaches the identity-channel capacity with energy constraint $k$:
 
\begin{align}
\frac{R^{\rm th}}{m}&=g(E/m)-\int_{0}^{\infty}ds \,q(s)H({\bf P}^{(s)})\nonumber\\&\geq g(k)-\frac{1}{2m }\log 2\pi e \left(k m+\frac{1}{12}\right)\nonumber\\&=C(\Phi,k) +O\left(\frac{\log m}{m}\right).\end{align}

\subsubsection{Low-energy regime}\label{lowbound}
Let us now discuss the low-energy regime for coherent-state strategies. The series that appears in the bound Eq.~\eqref{eq:uBound} can be rearranged as a power series in $E$, and at the leading order (at fixed $m$) it reads $\sum_{n=0}^{\infty} {\bf P}^{(E)}(n) \log\binom{n+m-1}{m-1}=E\log m +o(E) $. 
Hence the full low-energy expansion of the upper bound \eqq{eq:uBound} reads}
\be
R_{\rm c}^{\rm up}(\Phi_{m},E)= E\log\frac 1E +E(c-\gamma+\log m)+o(E),
\ee
where $c=\frac{e^{\frac 1 2+\gamma}}{2\sqrt{2}}-1\approx0.038$, and $\gamma$ is the Euler-Mascheroni constant. 

However, at extremely low energies ($10^{-60}\div10^{-40}$) and for all $m$, one can use the alternative bound $\tilde R_{\rm c}^{\rm up}(\Phi_{m},E)$,  
which provides the following asymptotically-tighter upper bound for the rate of our channel:
\begin{align}\label{eq:UBW}
\tilde R_{\rm c}^{\rm up}(\Phi_{m},E)&= E\log\frac 1E -E\log\log\frac 1E\nonumber\\&+E(2+\log 13-\gamma+\log m)+o(E).
\end{align}

An achievable lower bound is instead provided by a corresponding one for the Poisson channel. In the following we will adapt an on/off modulation that attains the Poisson channel capacity with unconstrained decoding at the leading order in $E$ and in general provides a good lower bound for $E\lesssim1$~\cite{Wang2014}. Note that this bound can be surpassed at larger energies by that of Ref.~\cite{Martinez2007}. The strategy we consider is a randomized on/off modulation (ROOM) and consists in sending a vacuum pulse $\ket{0}$ with probability $1-p$ and otherwise a Haar-random coherent pulse $\hat U\ket{\mathbf s}\otimes\ket{0}^{\otimes m-1}$ of energy $\abs{\mathbf s}^{2}=\frac Ep$. Following the same reasoning to obtain the optimal coherent-state rate, it is straightforward to see that the net effect of this encoding is that of inducing an on/off total energy distribution in \eqq{eq:maxRate}, i.e., $\{q(0)=1-p,q(\frac Ep)=p\}$. The corresponding rate is
{\small\be\begin{aligned}\label{eq:lowB}
R(\Phi_{m},E,p)&:=p\sum_{n=1}^{\infty} {\bf P}^{\left(E/p\right)}(n)\log\binom{n+m-1}{m-1}\nonumber\\&+h\left(1-p+p {\bf P}^{\left(E/p\right)}(0)\right)\\&+\sum_{n=1}^{\infty}h\left(p {\bf P}^{\left(E/p\right)}(n)\right)-p\sum_{n=0}^{\infty}h\left({\bf P}^{\left(E/p\right)}(n)\right)\\
&=p\sum_{n=1}^{\infty} {\bf P}^{\left(E/p\right)}(n)\log\binom{n+m-1}{m-1}\nonumber\\&+h\left(1-p+p{\bf P}^{\left(E/p\right)}(0)\right)\nonumber\\&+\left(1-{\bf P}^{\left(E/p\right)}(0)\right)h(p)-p h\left({\bf P}^{\left(E/p\right)}(0)\right),
\end{aligned}\ee}%
where we have defined $h(x)=-x\log x$ and used the property $h(x y)=x h(y)+y h(x)$. One can then maximize this function over $p\in[0,1]$ to obtain the best lower bound $R_{\rm c}^{\rm room}(\Phi_{m},E):=\max_{p}R(\Phi_{m},E,p)$. 

{We have already seen that, independently of the value of $p$, we have $p\sum_{n=1}^{\infty} {\bf P}^{\left(E/p\right)}(n)\log\binom{n+m-1}{m-1}\leq E\log m + o(E)$. Then we use the fact that for the remaining terms in Eq.~\eqref{eq:lowB}, at low energies, the maximum is attained for $p=E\log\frac 1E$, see~\cite{Wang2014}. By inserting this value of $p$ we get $p\sum_{n=1}^{\infty} {\bf P}^{\left(E/p\right)}(n)\log\binom{n+m-1}{m-1}=E\log m + o(E)$  
and therefore 
\be 
\small R_{\rm c}^{\rm room}(\Phi_{m},E)= E\log\frac 1E -E\log\log\frac 1E+E\log m+o(E).
\ee
Moreover, note that at low energies one can attain this rate at order $O(E)$ by explicit on/off modulation plus photodetection (OOP) that sends independently on each mode a fixed coherent pulse of energy $\frac E {pm}$ with probability $p$ and the vacuum otherwise~\cite{Wang2014} (see also~\cite{Vourdas1994} for a less-performing generalized PPM strategy). The rate for this strategy is immediately obtained from the on/off rate for the case $m=1$~\cite{Wang2014}:  
\begin{align}
R^{\rm oop}_{\rm c}(\Phi_m,E)&=m R^{\rm oop}_{\rm c}(\Phi_1,E/m)= E\log\frac 1E \nonumber\\&-E\log\log\frac 1E+E\log m+o(E).
\end{align}  
}

We can summarize the low energy analysis with the following: 
\begin{proposition}\label{upperbound} 
The rate of transmission of classical information through the channel $\Phi_m$ using low-energy coherent state encodings is bounded by 
\begin{align}
R_{\rm c}(\Phi_{m},E)&\geq E\log\frac 1E-E\log\log\frac 1E+ E\log m+o(E) \label{eq:lbLowE}\\
R_{\rm c}(\Phi_{m},E)&\leq E\log\frac 1E-E\log\log\frac 1E\nonumber \\&+ E(2+\log 13-\gamma+\log m)+o(E).
\end{align}
\end{proposition}

\subsection{Comparison of all strategies and squeezing advantage}

\begin{figure}[t!]
\includegraphics[scale=.4]{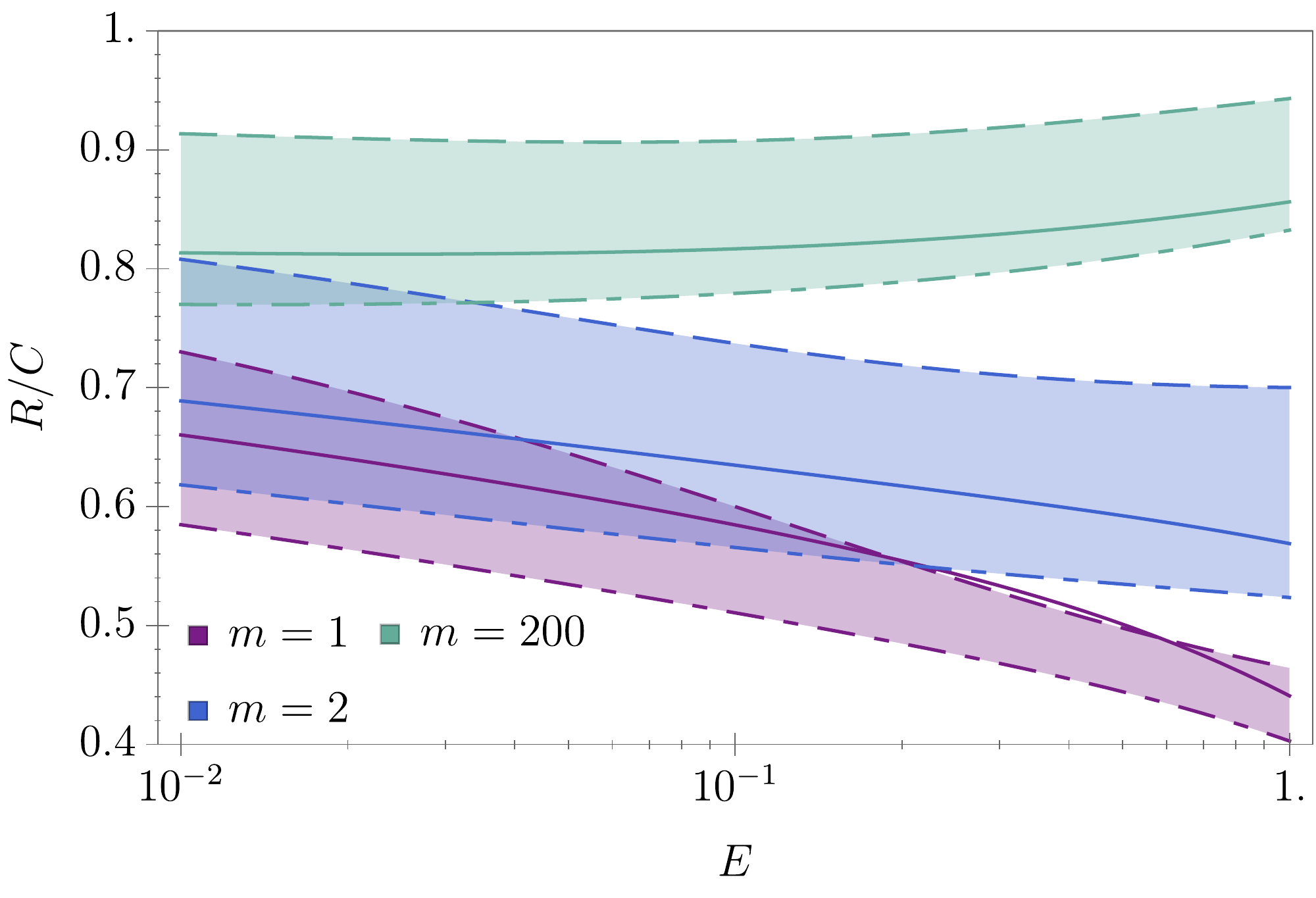} 
\caption{Plot (log-linear scale) of several rates per unit of channel capacity $C(\Phi_{m},E)$ vs. the average energy $E$ for several values of $m$: upper (dashed lines) and lower (dot-dashed lines) bounds on the optimal coherent-state rate, $R_{\rm c}^{\rm up/room}(\Phi_{m},E)$, lower bound (continuous line) on the optimal squeezed-coherent-state rate, $R_{\rm sc}^{\rm room}(\Phi_{m},E)$. The optimal { rate achievable with coherent states {rate}} lies in the shaded region. Note that as $m$ increases, the coherent-state rate becomes comparable with the capacity. Moreover, squeezing always provides a notable advantage over simple coherent encoding and it can even surpass the coherent-state encoding upper bound for $m=1$.}
\label{fig1}
\end{figure}

 \subsubsection{Quantum advantage via squeezing}
In Fig.~\ref{fig1} we plot the coherent and squeezed-coherent lower bounds $R^{\rm room}_{\rm c/sc}$ and the coherent upper bound $R^{\rm up}_{\rm c}$ per unit of channel capacity for several values of $m$, as a function of the mean energy $E$. First of all, we observe a general increasing trend of the rate with $m$, implying that one can make use of correlations in the phase noise to increase the Gaussian communication rate.

More importantly, the plot also shows that the use of squeezing provides a large increase of the communication rate for all $m$ and $E$ with respect to its direct coherent counterpart. This is particularly evident for $m=1$, where there even exists a small range of energies such that $R^{\rm room}_{\rm sc}(\Phi_{1},E)> R_{\rm c}^{\rm up}(\Phi_{1},E)$. Since for $m=1$ it also holds that $R^{\rm room}_{\rm sc}(\Phi_{1},E)=R^{\rm oop}_{\rm sc}(\Phi_{1},E)$ (see Sec.~\ref{subsubsec:oopRoom}), we conclude that the fully explicit OOP squeezed-coherent communication scheme surpasses any coherent-state scheme, answering a question that can be traced back to~\cite{Saleh1987}. This proves the existence of an unconditional quantum advantage with respect to classical communication strategies, which can be demonstrated using only experimental-friendly resources such as squeezing and photodetectors.

\begin{figure}[t!]
\includegraphics[scale=.4]{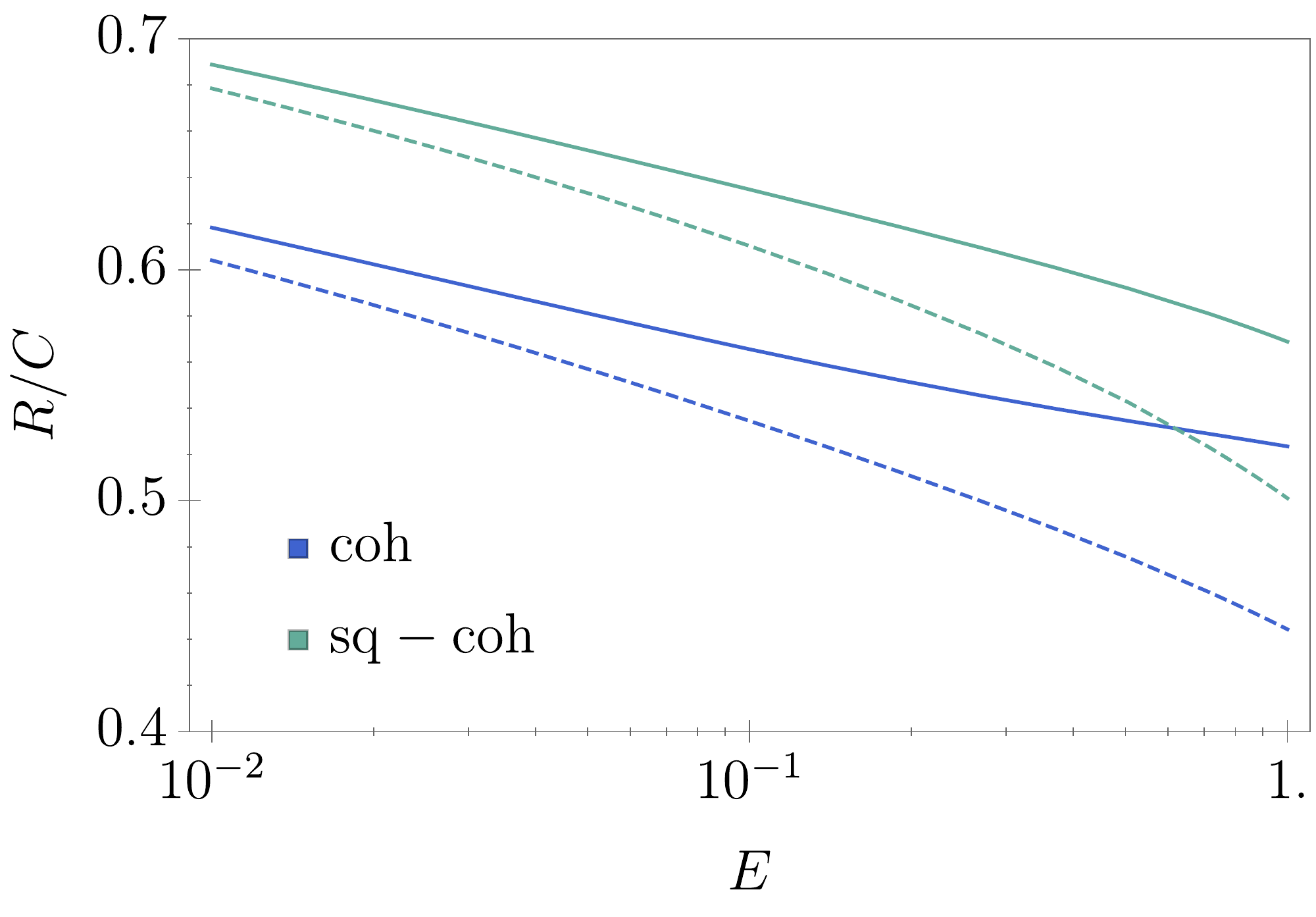}
\caption{Plot (log-linear scale) of several rates per unity of channel capacity: coherent and squeezed-coherent ROOM rate (continuous) and explicit OOP scheme rate (dashed), for $m=2$.} \label{fig:oopRoom}
\end{figure}

\subsubsection{Attainability of ROOM rate with fully explicity OOP scheme} \label{subsubsec:oopRoom}
We start by observing that, for $m=1$, the OOP scheme attains the lower bound \eqq{eq:lowBGen} at all energies both for coherent and squeezed-coherent encodings, i.e., $R^{\rm room}_{\rm c/sc}(\Phi_{1},E)=R^{\rm oop}_{\rm c/sc}(\Phi_{1},E)$, implying that both $(m=1)$-ROOM rates are attainable with an end-to-end explicit protocol. 

For $m>1$ instead we have $R^{\rm room}_{\rm c/sc}(\Phi_{1},E)>R^{\rm oop}_{\rm c/sc}(\Phi_{1},E)$ in general (see e.g. Fig.~\ref{fig:oopRoom}). 
Interestingly, in the low-energy regime this gap closes and the same lower bound of~\eqref{eq:lbLowE} can be attained by a fully explicit OOP strategy for all $m$. 
  
\subsubsection{Comparison with Fock and ternary encodings} 
The advantage of a ROOM squeezed-coherent encoding is quite small with respect to that obtained with an on/off encoding using the vacuum and a one-photon Fock state $\ket1$, which is, admittedly, not particularly challenging to produce nowadays. For this reason, we are driven to consider discrete encodings comprising more than two signals.

\begin{figure}[t!]\center
\includegraphics[scale=.4]{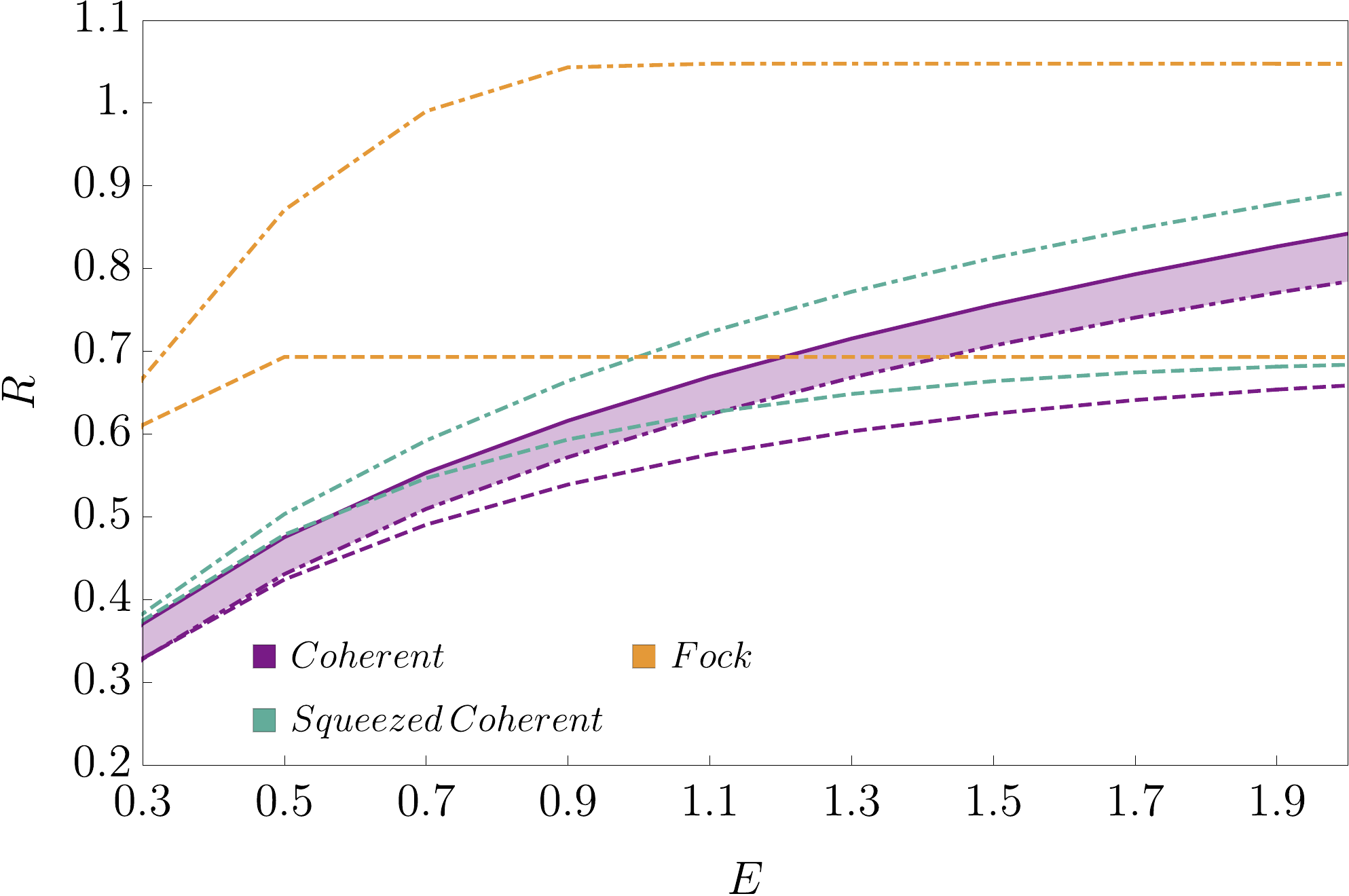} 
\caption{$m=1$, Binary (dashed) and ternary (dot-dashed) rates achievable with Fock encodings with up to two photons (orange), coherent states (violet), and squeezed coherent states (green). The violet solid line is the upper bound on the coherent state rate Eq.~\ref{cohlossupper}}
\label{fig:ternary}
\end{figure}

As we already pointed out, for any ensembles of $t$ states the maximum rate is bounded by $\log t$. For restricted Fock ensembles with photon number up to $1$ or $2$ the Holevo quantity rapidly saturates to respectively $\log 2$ and $\log 3$. Still, as shown in Fig.~\ref{fig:ternary}, the advantage of squeezed-coherent encodings is enhanced by considering ternary constellations, which can surpass the performance of $0/1$ Fock encodings. Importantly, the amount of squeezing required by these optimal encodings is relatively modest, e.g., at $E=1.1$ the largest $r\simeq0.62$ corresponds to a squeezing of $5.4\rm dB$, while at $E=2$ the largest $r\simeq0.68$ corresponds to a squeezing of $5.8\rm dB$. Hence our squeezed-coherent encoding is fully within reach of current experimental platforms~\cite{Vahlbruch2016} (up to $15\rm dB$), even for on-chip production~\cite{Zhang2021}(up to $8\rm dB$).

Clearly, one can surpass the ternary squeezed-coherent encoding with a ternary Fock encoding, as shown in Fig.~\ref{fig:ternary}. However, the difficulty of producing Fock states with photon number larger than $1$, makes encodings with multiple squeezed-coherent states preferable over those relying on Fock states.
Such advantage is further enhanced in the presence of loss, as we will discuss in the next section.

\section{Beating Fock encodings under loss}\label{sec:loss}

A simple model to account for losses is to precede the phase noise channel $\Phi_m$ with a lossy bosonic channel $\mathcal E_{\eta, n_{th}}^{\otimes m}$. $\mathcal E_{\eta, n_{th}}$ sends Gaussian states into Gaussian states, acting on first and second moments as
 \begin{align}
&\boldsymbol\mu\xrightarrow{\mathcal{E}_{\eta,n_{th}}} \bm\mu'=\sqrt{\eta}{\bm \mu}\;,\\
&V\xrightarrow{\mathcal{E}_{\eta,n_{th}}}V'=\eta V + (1-\eta)(2n_{th}+1)I_2\,  ,
\end{align}

and for $n_{th}=0$ it acts on Fock states as following

\begin{equation}\label{Fockloss}
\mathcal E_{\eta, 0}(\ketbra{n}{n})=\sum_{i=0}^n \binom{n}{i}\eta^i(1-\eta)^{n-i}\ketbra{i}{i}
\end{equation}

The action on the characteristic function of $\Phi_m$ and $\mathcal E_{\eta, n_{th}}^{\otimes m}$ is
\begin{align}
&\phi(\mathbf r)\xrightarrow{\mathcal{E}_{\eta,n_{th}}} \phi'(\mathbf r)= \phi(\eta \mathbf r)e^{-\frac{(1-\eta)(2n_{th}+1)}{4}|\mathbf r|^2}\;,\\
&\phi(\mathbf r)\xrightarrow{\Phi_m} \phi'(\mathbf r)=\frac{1}{2\pi} \int_0^{2\pi} d\theta \phi({R(\theta)\oplus\cdots\oplus R(\theta)}\mathbf r)\;,
\end{align}
where each $R(\theta)$ is a rotation of angle $\theta$ of the coordinates $(x_i,p_i)$. From this characterization it is evident that $\Phi_m$ and $\mathcal E_{\eta, n_{th}}^{\otimes m}$ commute, therefore the order in which we place them is irrelevant.

Replacing ensembles $\cE=\{q(x),\hat\rho^{(x)}\}$ with $\cE_\eta=\{q(x),\mathcal E_{\eta, n_{th}}^{\otimes m}(\hat\rho^{(x)})\}$ in the rate expression Eq.~\ref{eq:maxRateGen} we can compute rates in the presence of loss. 

Since the action of loss on coherent states is $\mathcal{E}_{\eta,0}(\ketbra{\alpha}{\alpha})=\ketbra{\sqrt{\eta}\alpha}{\sqrt{\eta}\alpha}$, we can immediately compute the maximum coherent-state rate in the presence of loss as 
\begin{equation}\label{cohlossupper}
R_c(\Phi_m\circ \mathcal{E}_{\eta,0},E) = R_c(\Phi_m, \eta E),
\end{equation} 
and employ upper and lower bounds adapted from Sec.~\ref{sec:bounds}.

To compute the rates of encodings generated by applying a random PI to single-mode squeezed-coherent states, we need instead the photon-number distribution of a generic Gaussian state, which is reported in Appendix~\ref{photG}, Eq.~(\ref{thermphot}). Finally, to compute the rates of encodings generated by Fock states we can use Eq.~(\ref{Fockloss}).

\begin{figure}[t!]\center
\includegraphics[scale=.4]{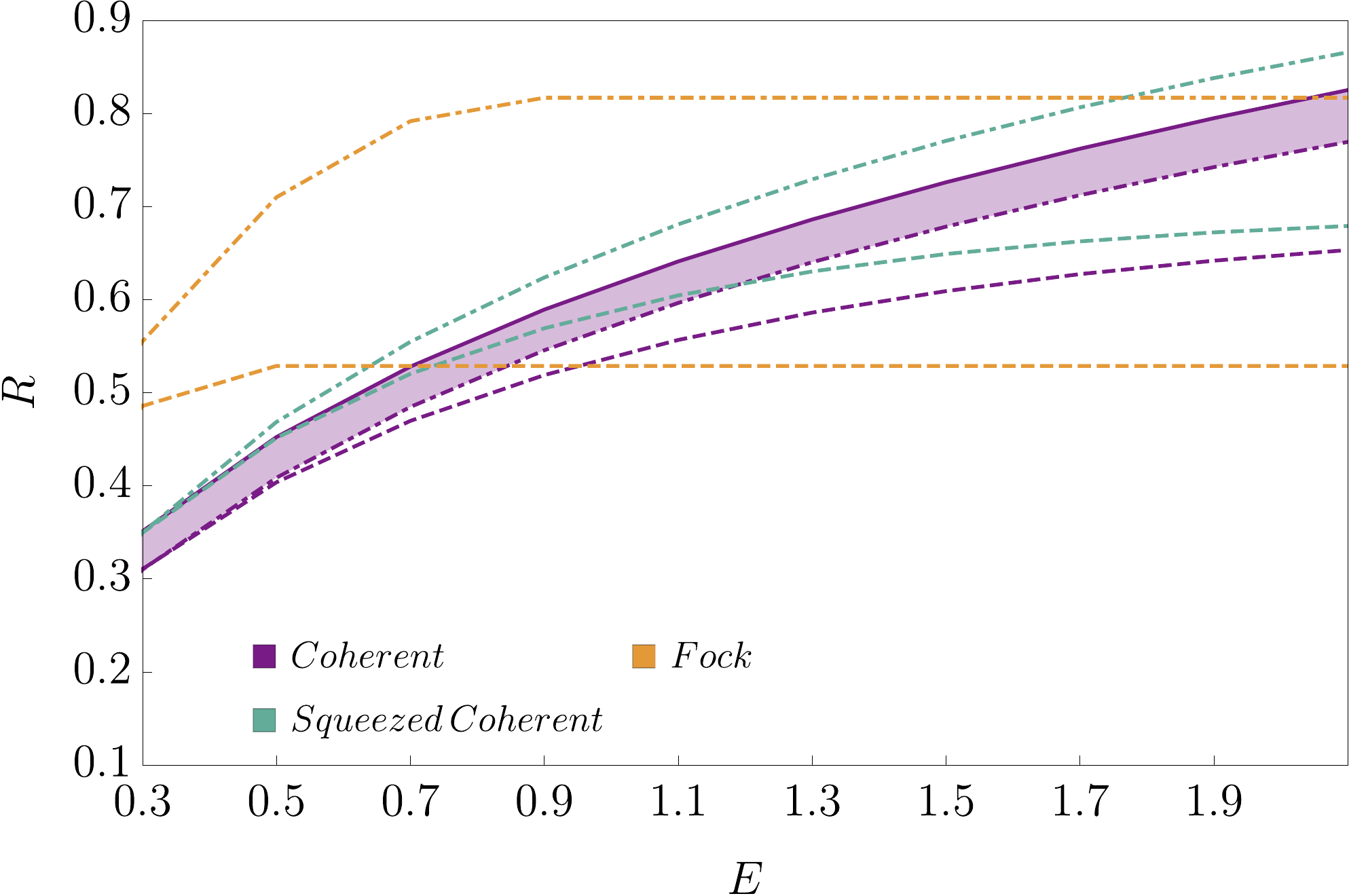} $\eta=0.9$
\includegraphics[scale=.4]{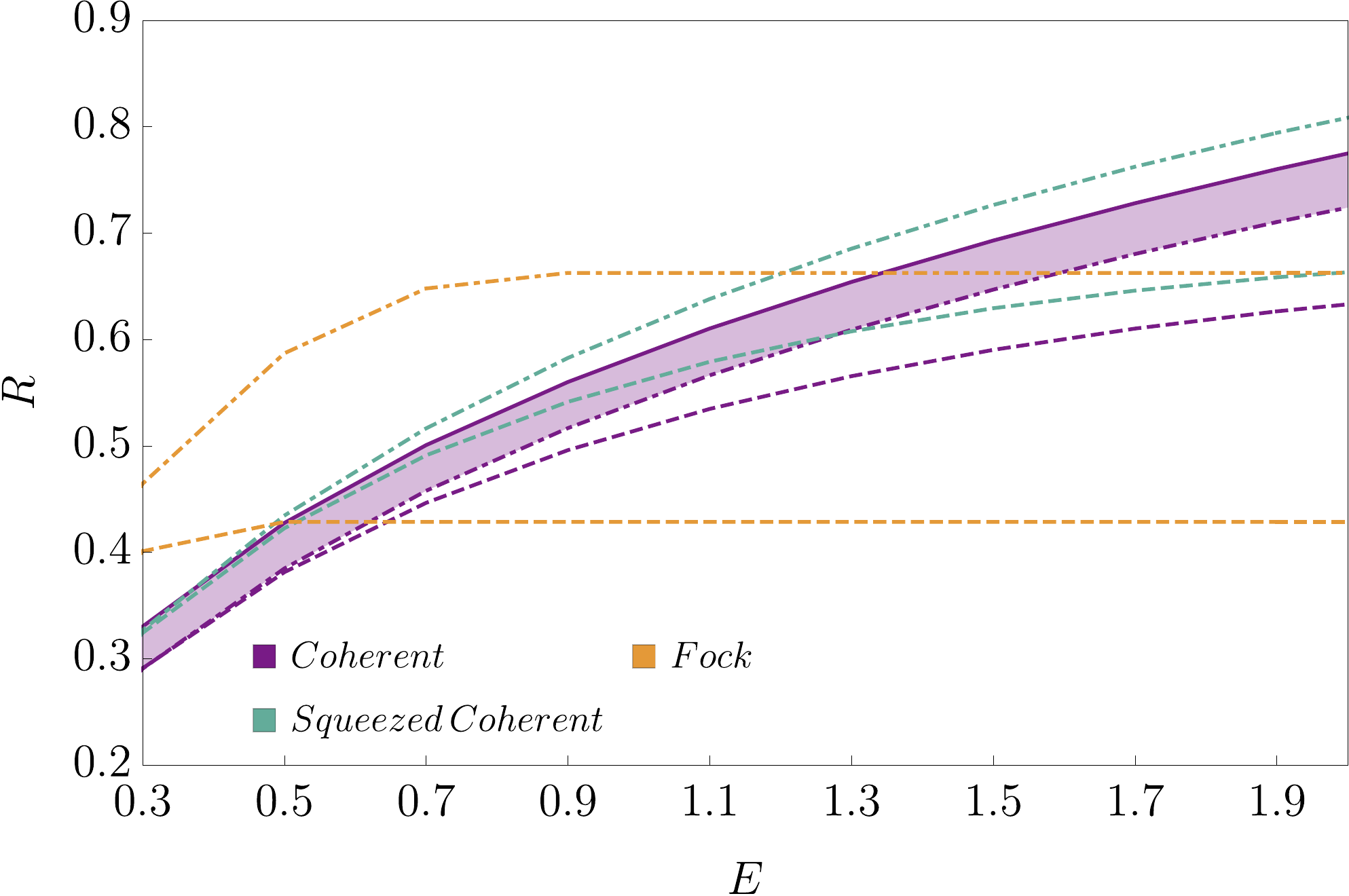} $\eta=0.8$
\includegraphics[scale=.41]{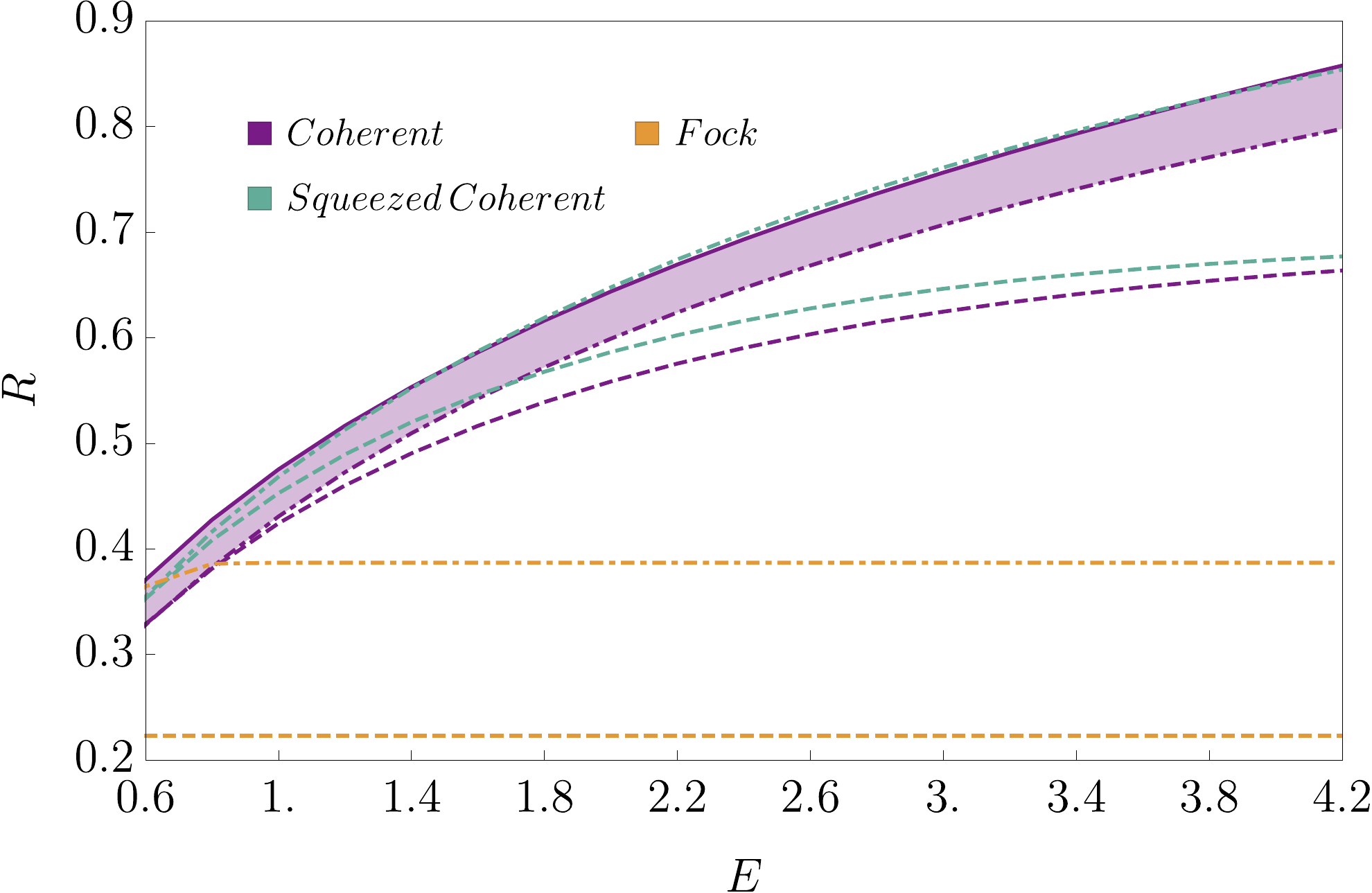} $\eta=0.5$
\caption{$m=1$, Binary (dashed) and ternary (dot-dashed) rates achievable with Fock encodings with up to three photons (orange), coherent states (violet), and squeezed coherent states (green). The violet solid line is the upper bound on the coherent state rate Eq.~\ref{cohlossupper}.}
\label{figloss}
\end{figure}

\begin{figure}[t!]\center
\includegraphics[scale=.4]{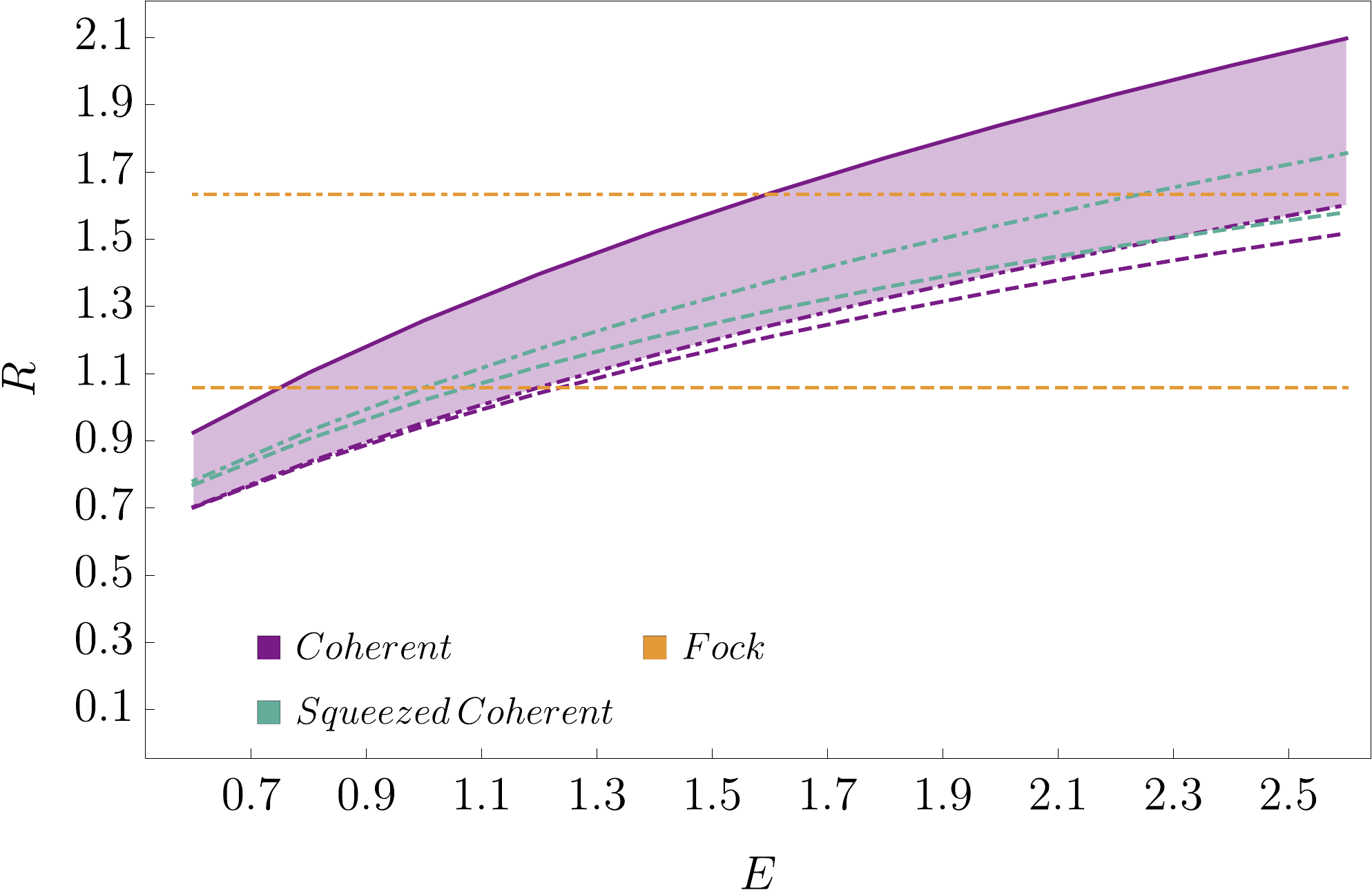} $\eta=0.9$
\caption{$m=2$, upper bounds on rate with Fock encodings with up to three photons (removing energy constraint), randomized binary (dashed) and randomized ternary (dot-dashed) rates achievable with coherent states (violet), and squeezed coherent states (green). The violet solid line is the upper bound on the coherent state rate Eq.~\ref{cohlossupper} }
\label{figloss2}
\end{figure}

In Figs.~\ref{figloss},\ref{figloss2} we plot these rates in the case $m=1$ and $m=2$ and zero-temperature environment, $n_{th}=0$, showing that squeezed-coherent encodings outperform both coherent and Fock encodings with photon number up to $2$ in the presence of a moderate amount of loss, in the regime $E\approx1$.

\section{Communication cost of establishing a phase reference}\label{sec:phaseRef}
\begin{figure}[t!]
\includegraphics[scale=.4]{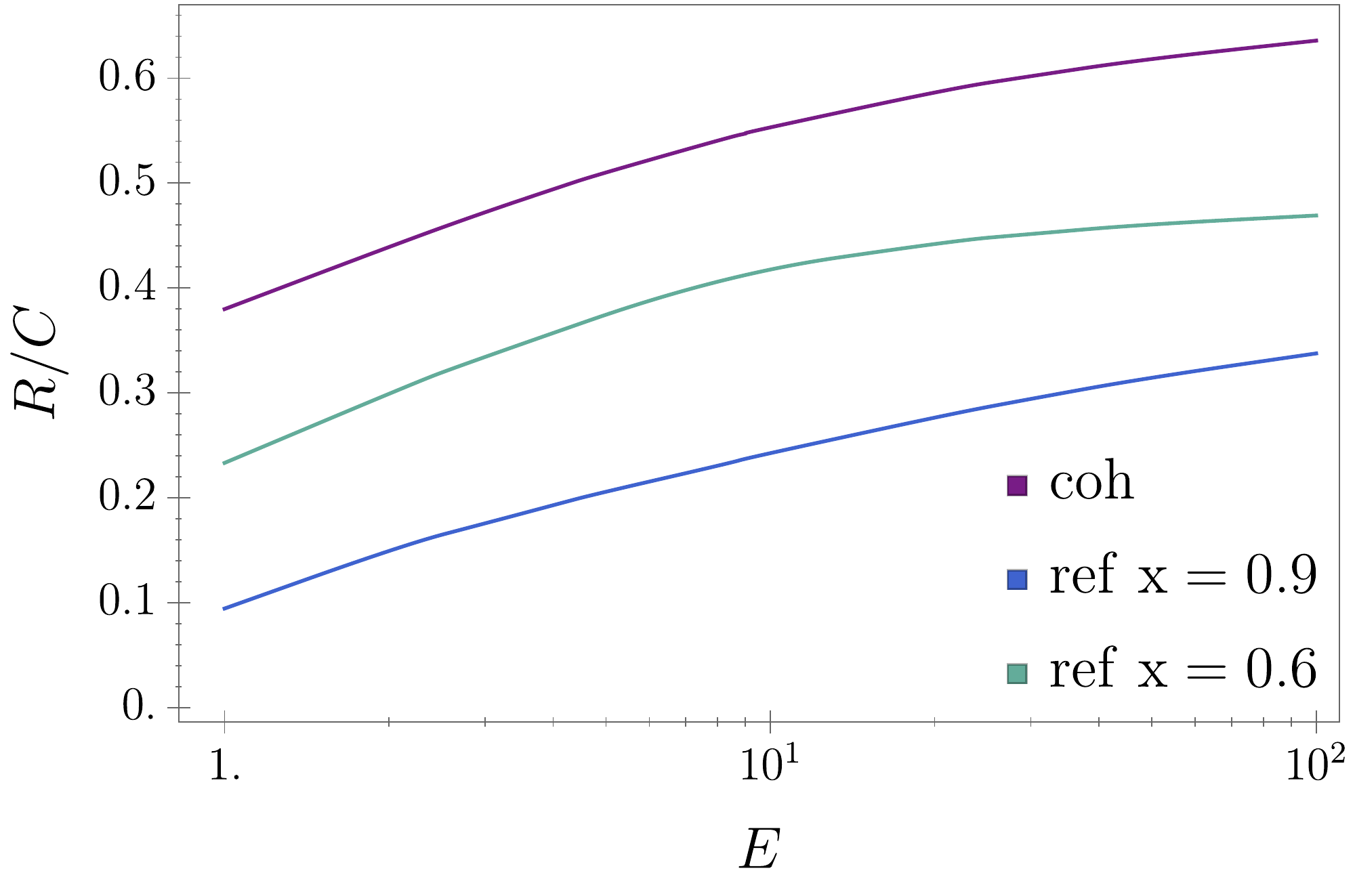}
\caption{Plot (log-linear scale) of several rates per unity of channel capacity: Gaussian coherent-state ensembles on all the $m=2$ modes or with a fixed reference on one mode.} \label{fig:refFrame}
\end{figure}
The optimality of covariant encodings makes strategies with phase reference states suboptimal in principle, but it is still worth to compare them with covariant encodings. We consider an encoding employing $Ex$ energy in the first mode to prepare a fixed phase reference state and $E(1-x)$ on the remaining $m-1$ modes with arbitrary encoding, with $0<x<1$. By monotonicity of the capacity, clearly this rate cannot be better than that of the identity channel on $m-1$ modes with energy constraint $E(1-x)$. Asymptotically at high energy, the leading term of this upper bound is $(m-1)\log E$, independently of $x$ and of the reference state, which is less than the coherent-state rate achieving Eq.~(\ref{highen}) by $\frac 1 2 \log E$. We give an expression of the rate with phase synchronization in Appendix~\ref{phasesynch}, showing that coherent states encoding achieve this upper bound asymptotically. In Fig.~\ref{fig:refFrame} we show the comparison in the finite energy regime between a covariant coherent encoding with an average thermal input state and encodings using a truncated phase state $\ket{\psi}=[2xE+1]^{-1/2}\sum_{n=0}^{2xE}\ket{n}$ as reference and a thermal ensemble of coherent states for coding. In fact, while phase estimation procedures~\cite{Caves1981,Monras2006a,Yonezawa2012,Safranek2014,Sparaciari2015} {{where improvements are} which benefit} from super-Poissonian photon-number statistics, the advantage we report in this paper
is obtained by trading signals characterized by Poissonian photon-number distribution with sub-Poissionian squeezed-coherent states.

\section{Discussion and conclusions}\label{sec:conclusions}
We have analyzed the performance of Gaussian encodings in the presence of phase-noise with a finite decoherence time, such that $m$ successive signals can be sent before losing the phase reference. This is a physically-motivated example of non-Gaussian channel, and we showed that good encodings make an intelligent use of the relative degrees of freedom, rather than trying to synchronize a common phase. Indeed, phase synchronization schemes with quantum-enhanced phase estimation seem to be unfavored with respect to general coherent-state strategies, if the global energy cost is taken into account. 

Moreover, we showed that squeezing can greatly enhance the communication rate, as an effect of reducing the entropy of the total-photon-number distribution. In particular for $m=1$ we proved that, for the first time to our knowledge, an explicit strategy, alternating between the vacuum and a squeezed-coherent state, together with photodetection, outperforms any coherent-state code. This is particularly interesting considering that it can be easily realized with current technology, while the as-yet-unknown optimal coherent-state rate will need in general the use of entangled measurements at the receiver side, which are still challenging.

Finally, we showed that the squeezing advantage over coherent states is robust with respect to additional loss effects in the communication line and that, in this case, squeezed-coherent encodings with multiple pulses can even outperform Fock-state encodings. This fact, in conjunction with the difficulty of realizing photon-number states and the relatively small amount of squeezing required by our strategies, establishes squeezed-coherent states as a robust and efficient coding method for communication without phase-reference.

We leave as open questions: the optimality of strategies employing non-zero squeezing among Gaussian states for any $m$ and $E$; the sub-optimality of ensembles using states with super-Poissonian statistics, which is good for phase synchronization, with respect to coherent-state strategies. Moreover, we did not consider the possibility of sending entangled squeezed states across the channel uses, which could in principle further enhance the communication rates due to superadditivity.

\section{Acknowledgments}
M.\,F. and V.\,G. acknowledge support by MIUR via PRIN 2017 (Progetto di Ricerca di Interesse Nazionale): project QUSHIP (2017SRNBRK). 
M.\,R., M.\,S. and J.\,C. acknowledge support from the Spanish MINECO, project FIS2016-80681-P with the support of AEI/FEDER funds; the Generalitat de Catalunya, project CIRIT 2017-SGR-1127. This project has received funding from the European Union's Horizon 2020 research and innovation programme under the Marie Sk\l odowska-Curie grant agreement No 845255.
M.\,S. also acknowledges support from the Baidu-UAB collaborative project ``Learning of Quantum Hidden Markov Models''.


\appendix

\section{Decomposition into irreducible representations of $\mathrm U (m)$}\label{coherentgroups}
In this section we determine the decomposition into irreducible representations of $\mathrm U (m)$ of the Hilbert space of $m$ modes. In the following we switch to a complex notation for coherent states, i.e., $\ket\alpha:=e^{\alpha \hat a^{\dag}-\alpha^{*}\hat a}\ket 0$. 

Since coherent states are an overcomplete set, we can first restrict to study the action of $\mathrm {U} (m)$ on coherent states {and then straightforwardly extend the result to arbitrary bosonic states via the decomposition 
\be\label{eq:Prep}
\hat\rho=\int d^{2m}\vec\alpha~{\cal P}_{\rho}(\vec\alpha)\dketbra{\vec\alpha},
\ee
where ${\cal P}_{\rho}(\vec\alpha)$ is the Glauber-Sudarshan $P$-representation~\cite{glauber,serafiniBOOK} of the $m$-mode bosonic state $\hat\rho$.}

We will make use of a crucial property that connects coherent states of an infinite-dimensional system with spin-coherent states of finite dimension~\cite{Perelomov1972,Zhang1990}. First, note that an m-mode coherent state can be decomposed as
 \begin{equation}
\ket{\vec\alpha}=\sum_{n=0}^{\infty}\sqrt{{\bf P}^{(s)}(n)} \ket{\psi_{n}(\vec \alpha)},
 \end{equation}
where $s:=|\vec\alpha|^2$ the mean energy of the state, and $\hat\Pi_n\ket{\vec \alpha}= \sqrt{{\bf P}^{(s)}(n)} \ket{\psi_n(\vec\alpha)}$. Explicitly
\begin{align}\label{repc}
\ket{\psi_n(\vec\alpha)}
&=\sum_{\sum_{i=1}^{m} n_i=n}  \sqrt{\binom{n}{\{n_i\}}} \prod_{i=1}^{m} u_i^{n_i} \ket{\vec n}\nonumber
\end{align}
where we have introduced the multi-mode Fock state $\ket{\vec n}=\ket{n_{1}}\otimes\cdots\otimes\ket{n_{m}}$, and $\vec u:=\frac{\vec \alpha}{|\vec \alpha|}$. 
Now observe that each $\ket{\psi_{n}(\vec \alpha)}$ lives in a finite-dimensional subspace and it can be mapped to the state of $n$ copies of a $m$-level system state with coefficients $\vec u$: 
\begin{align}\small
(\sum_{i=1}^{m}u_i \ket{i})^{\otimes n}&= \sum_{\sum_{i=1}^{m} n_i=n}  \prod_{i=1}^m u_{i}^{n_i} \sum_{\sigma \in S_n} U(\sigma) \ket{\vec n^{(m)}}\nonumber\\&\cong\ket{\psi_n(\vec u)},
\end{align}
where $\ket{\vec n^{(m)}}$ is the tensor-product state $\ket{\vec n^{(m)}}=|\underbrace{1,\cdots,1}_{n_1},\cdots \underbrace{m,\cdots,m}_{n_m}\rangle$, with $n_{i}$ repetitions of the $i$-th basis element, $U(\sigma)$ is a permutation of the $m$-level systems and the isomorphism is defined on the basis of permutation-symmetric states $\binom{n}{\{n_i\}}^{-1/2}\sum_{\sigma \in S_n} U(\sigma) \ket{\vec n^{(m)}}\rightarrow\ket{\vec n}$.
Finally, thanks to this mapping, the action of an energy-preserving Gaussian unitary $\hat{U}$ corresponding to $U\in {\rm U}(m)$ in phase space, can also be written as
\begin{equation}\label{repu}
\hat U \ket{\vec \alpha}= \ket{ U \vec \alpha}=\sum_{n=0}^{\infty} \sqrt{{\bf P}^{(s)}(n)} \hat d^{(n,m)}_{U} \ket{\psi_n(\vec u)},
\end{equation}
where $\hat d^{(n,m)}_{U}$ is the image of $U$ with respect to the irreducible representation of ${\rm {\rm U}}(m)$ on the permutation-symmetric subspace of $n$ $m$-level systems. This is enough to conclude that each block with total photon number $n$ hosts the irreducible representation of $\mathrm U (m)$ corresponding to the Young diagram of one row of length $n$, which has dimension $\binom{n+m-1}{m-1}$\cite{Hayashi}. 

{By Schur's lemma it then follows that the Haar average decoheres blocks with different total photon numbers and, inside each block with fixed total photon number, it acts as a ${\rm {\rm U}}(m)$-twirling:
\begin{align}
&\int_{{\rm U}(m)}dU \hat U \dketbra{\vec \alpha}\hat U^{\dag}\nonumber\\& = \sum_{n=0}^{\infty} {\bf P}^{(s)}(n) \int_{{\rm U}(m)} dU \hat d^{(n,m)}_{U} \dketbra{\psi_n(\vec u)} \hat d^{(n,m)\dag}_{U}\nonumber\\&=\sum_{n=0}^{\infty} {\bf P}^{(s)}(n) \frac{\hat \Pi_{n}}{\binom{n+m-1}{m-1}}.
\end{align}
This result can then be applied to each coherent-state term in the decomposition of Eq.~\eqref{eq:Prep}, obtaining Eq.~\eqref{eq:twirling} 
 of the main text. }

{\section{Pure-state ensembles are always optimal among Gaussian encodings}\label{puregauss}
Consider an ensemble comprising general Gaussian states of the form $\hat\rho_{\rm G}=\hat U \hat S(\mathbf{r}) \hat D(\mathbf{s})\hat\rho_{\rm th}\hat D^{\dag}(\mathbf{s})\hat S^{\dag}(\mathbf{r})\hat U^{\dag}$, where $\hat\rho_{\rm th}$ is an $m$-mode thermal state, $\hat U$ is an $m$-mode PI and $\hat D(\mathbf{s})$, $\hat S(\mathbf{r})$ are {as defined in Eq.~(\ref{eq:dispacement}) and~(\ref{eq:operations}). {the tensor product of single-mode displacement operators $\hat D(\_{i})=\exp(\alpha_{i} \hat a_{i}^{\dag}-\alpha_{i}^{*}\hat a_{i})$ and squeezing operators $\hat S(r_{i})=\exp(\frac {r_{i}}{2}(\hat a_{i}^{2}-\hat a_{i}^{\dag 2}))$, respectively.}} 
Now recall that any thermal state can be decomposed as a mixture of coherent states with Gaussian weights, i.e., $\hat\rho_{\rm th}=\int d^{2m}\vec\beta~p_{\rm G}(\vec\beta) |\vec\beta\rangle\langle\vec\beta|$~\cite{serafiniBOOK} and hence every Gaussian state $\hat\rho_{\rm G}$ can be written as a mixture of pure Gaussian states with Gaussian weight. 
Then for any mixed-state Gaussian ensemble $\cE_{\rm G}:=\{q(x),\hat\rho_{\rm G}(x)\}$, respecting the mean-energy constraint, one can consider an equivalent pure-state Gaussian ensemble $\tilde\cE_{\rm G}:=\{q(x) p_{\rm G}(\vec\beta|x), \hat\Psi_{G}(\vec\beta,x)\}$, comprising all the pure states $\hat\Psi(\vec\beta,x)=\dketbra{\psi(\vec\beta,x)}$, with $\ket{\psi(\vec\beta,x)}=\hat U_{x} \hat S(\mathbf{r}_{x}) \hat D(\mathbf{s}_{x})\ket{\vec\beta}$, that take part in the decomposition of some $\hat\rho_{\rm G}(x)$, with proper weights. Then by the equivalence of these two ensembles and the concavity of the entropy we obtain, for any channel $\Phi$ acting on $m$ bosonic modes, 
\be\begin{aligned}
\chi(\Phi,\cE_{\rm G})&=S\left(\int dx\, q(x)\Phi(\hat\rho_{\rm G}(x))\right)\\&-\int dx\, q(x)S\left(\Phi(\hat\rho_{\rm G}(x))\right)\\
&\leq S\left(\int dx\,d^{2m}\vec\beta\, q(x)p_{\rm G}(\vec\beta|x)\Phi(\hat\Psi(\vec\beta,x))\right)\\&-\int dx\,d^{2m}\vec\beta\, q(x)p_{\rm G}(\vec\beta|x)S\left(\Phi(\hat\Psi_{}(\vec\beta,x))\right)\\&=\chi(\Phi,\tilde\cE_{\rm G}).
\end{aligned}\ee
This implies that, when optimizing the Holevo quantity over Gaussian encodings, one can always restrict to pure states.}

\section{Communicate with phase reference}\label{phasesynch}

Consider now the scenario where Alice and Bob use a fraction of the total available energy $x E$ to prepare a single 
mode state suitable for estimating the phase of the channel and $(1-x)E$ is the average energy of the ensemble of coherent states on the remaining $m-1$ modes. The input states have thus the form $\ket{\psi}\otimes \ket{\vec \alpha}$, with $\ket{\vec\alpha}=\otimes_{i=2}^{m}\ket{\alpha_i}$, $\bra{\psi}\hat n_{1}\ket{\psi}=xE$, $\bra{\vec \alpha}\hat \sum_{i=2}^{m}\hat n_{i}\ket{\vec \alpha}=|\alpha|^2$. Since $\Phi_m$ commutes with energy-preserving Gaussian unitaries on the last $m-1$ modes, one can adapt the argument in the main text to obtain an optimal rate

\begin{align}
&\chi_{\rm c}^{\rm ph}(\Phi_m, E,x)\nonumber\\&=\Bigg[S\left(\int dp(\vec \alpha)\Phi_m( \dketbra{\psi}\otimes \int dU \,\hat U\dketbra{\vec \alpha}\hat U)\right)\nonumber\\&-\int dp(\vec \alpha) \int dU S(\Phi_m( \dketbra{\psi}\otimes \hat U\dketbra{\vec \alpha}\hat U))\Bigg]	\nonumber\\&=
\Bigg[S\Big(\int dp(\vec \alpha)\Phi_m\Big( \dketbra{\psi}\otimes \sum_{n=0}^{\infty}{\bf P}^{\left(|\alpha|^2\right)}(n)\nonumber\\
&\cdot\frac{ \hat{\Pi}_n^{(m-1)}}{\binom{n+m-2}{m-2}}\Big)\Big)-\int dp(\vec \alpha) S(\Phi_m( \dketbra{\psi}\otimes \dketbra{\vec \alpha }))\Bigg].
\end{align}
where $\hat{\Pi}_n^{(m-1)}$ is the projector on the space of $m-1$ modes with total photon number $n$. The first term is
\begin{align}
&\Phi_m( \dketbra{\psi}\otimes \sum_{n=0}^{\infty}{\bf P}^{\left(|\alpha|^2\right)}(n)\frac{\hat{\Pi}_n^{(m-1)}}{\binom{n+m-2}{m-2}})
\nonumber\\&=\sum_{l=0}^{\infty} {\bf q}^{(xE)}(l) \dketbra{l}\otimes \sum_{n=0}^{\infty} {\bf P}^{\left(|\alpha|^2\right)}(n)\frac{\hat{\Pi}_n^{(m-1)}}{\binom{n+m-2}{m-2}},
\end{align}
where $ {\bf q}^{(xE)}(n):=\tr{\hat \Pi_n \dketbra{\psi}\otimes \dketbra{0}}$, and the second term can be computed by noting that $\tr{\hat \Pi_n \dketbra{\psi}\otimes \dketbra{\vec\alpha}}=\sum_{l=0}^{n}  {\bf q}^{(xE)}(l) {\bf P}^{\left(|\alpha|^2\right)}(n-l)$.
Therefore, denoting ${\bf P}^{(s,E)}$ the probability distributions such that ${\bf P}^{(s, x E)}(n)=\sum_{l=0}^{n}  {\bf q}^{(xE)}(l) {\bf P}^{\left(s\right)}(n-l)$, the rate is
\begin{align}
&\chi_{\rm c}^{\rm ph}(\Phi_m, E,x)=
S[{\bf q}^{(xE)}]+S[\int_{0}^{\infty} ds\, p(s)\, {\bf P}^{(s)}]\nonumber\\&+\sum_{n=0}^{\infty}\int_{0}^{\infty} ds\,p(s)\, {\bf P}^{(s)}(n)\log \binom{n+m-2}{m-2}\nonumber\\&-\int_{0}^{\infty} ds\,p(s)\,S[{\bf P}^{(s,xE)}]
\end{align}
Using a coherent state as reference, and coding with a thermal ensemble for $m-1$ modes, at high energies we obtain, for fixed $x$,  $0<x<1$,
\begin{align}
\chi_{\rm c}^{\rm ph}(\Phi_m, E,x)&=
(m-1)\log E+ \frac{1}{2}\log E\nonumber\\& - \frac{1}{2}\log E +O(1)
\end{align}
which is already sufficient to reach the upper bound at leading order. Other phase reference states are not useful at this level.

\section{Squeezed-coherent encodings}\label{app:sqCoh}
The photon-number distribution of a coherent squeezed state $\hat S(r)\hat D((\sqrt{2}\mathrm{Re}\,\alpha,\sqrt{2}\mathrm{Im}\,\alpha))\ket{0}$, $r\in\mathbb{R}$ and $\alpha\in\mathbb{C}$, is given by $p(n|r,\alpha)=\abs{c(n|r,\alpha)}^{2}$, where~\cite{Yuen1976,Gong1990}
\begin{align}
&c(n|r,\alpha)=(n ! \cosh(r))^{-\frac12}\left(\frac12 \tanh(r)\right)^{n / 2} \nonumber\\\times&H_{n}\left[\alpha\sinh(2r)^{-1 / 2}\right]\exp \left[-\frac{1}{2}|\alpha|^{2}-\frac12 \tanh(r) \alpha^{2}\right]
\end{align}
and $H_{n}(\gamma)$ is the Hermite polynomial of order $n$. Taking $\alpha\in\mathbb{R}$, the average energy of the state is $E+\frac12=\frac12\cosh(2r)+e^{-2r}\alpha^{2}$. Substituting for $\alpha$ we then obtain
\begin{align}\small
&c(n|r,E)=(n ! \cosh(r))^{-\frac12}\left(\frac12 \tanh(r)\right)^{n / 2}\nonumber\\&\times H_{n}\left[\sqrt{\frac{(2E+1-\cosh(2r))e^{2r}}{2\sinh(2r)}}\right] \nonumber\\&\times\exp \left[-\frac{(2E+1-\cosh(2r))e^{2r}}{4}(1+\tanh(r))\right].
\end{align}
An achievable rate using these states for the encoding is obtained via the following on/off modulation:
\begin{align}
&\cE_{p,\mathbf{r},\vec \alpha,U}=\left\{(1-p)\dketbra{0}^{\otimes m}, \right.\nonumber\\&\left. p\,dU\,\hat U \hat S(\mathbf{r}) \hat D(\vec{\alpha})\dketbra{0}^{\otimes m} \hat D(\vec{\alpha})^{\dag} \hat S(\mathbf{r})^{\dag} \hat U^{\dag}\right\},
\end{align}
where the vacuum state is sent with probability $(1-p)$, while a pulse is sent with probability $p$. The latter is generated by a product of displacements {denoted as $\hat D(\vec \alpha):=D((\sqrt{2}\mathrm{Re}\,\alpha_1,\sqrt{2}\mathrm{Im}\,\alpha_1))\cdots D((\sqrt{2}\mathrm{Re}\,\alpha_m,\sqrt{2}\mathrm{Im}\,\alpha_m))$,} and single-mode squeezing with fixed parameters on each mode, $\mathbf{r},\vec{\alpha}\in\mathbb{R}^{m}$, followed by a Haar-random passive Gaussian unitary $\hat U$ on the $m$ modes. All the parameters $p,\mathbf{r},\vec{\alpha}$ are chosen so as to satisfy an average-energy constraint for the ensemble and the total photon number distribution is ${\bf Q}^{(\mathbf{r},\vec\alpha)}$, with probabilities
\begin{equation}
{\bf Q}^{(\mathbf{r},\vec\alpha)}(n)=\sum_{\sum_{i}{n_i}=n }\prod_{i=1}^{m}p(n_{i}|r_{i},\alpha_{i}).
\end{equation}

Following the same reasoning leading to~\eqref{eq:lowB}, the rate achievable with this encoding is 
\be\begin{aligned}
&R(\Phi_{m},E,\vec \alpha,\mathbf{r},p)=p\sum_{n=1}^{\infty}{\bf Q}^{(\mathbf{r},\vec\alpha)}(n)\log\binom{n+m-1}{m-1}\\&+h\left(1-p+p{\bf Q}^{(\mathbf{r},\vec\alpha)}(0)\right)+\left(1-{\bf Q}^{(\mathbf{r},\vec\alpha)}(0)\right)h(p)\\&-p\,h\left({\bf Q}^{(\mathbf{r},\vec\alpha)}(0)\right).\label{eq:rateSqc}
\end{aligned}\ee

The rates of covariant ensembles generated from ternary encodings can be obtained by Eq.~(\ref{eq:maxRateGen}) simply putting ${\bf Q}^{(\mathbf{r},\vec\alpha)}(n)$ as total photon-number distribution. 

\section{Photon number distribution of single mode Gaussian states}\label{photG}
Single-mode Gaussian states $\hat \rho$ are characterized by the vector $\mathbf m=(x,p):=(\tr{\hat \rho \hat x},\tr{\hat \rho \hat p})$, where $\hat x$ and $\hat p$ are the position and momentum operator, and the matrix 
\begin{equation}
\small V:=\left(\begin{matrix} \tr{2(\hat x -x)^2\hat \rho} & \tr{\{(\hat x -x),(\hat p -p)\}\hat \rho}  \\ \tr{\{(\hat x -x),(\hat p -p)\}\hat \rho}  &  \tr{2(\hat p -p)^2\hat \rho} \end{matrix}  \right)
\end{equation}

Their photon number distribution can be computed as follows~\cite{xu2010photon}. Redefining $\hat x$ and $\hat p$ with a rotation in the phase space, $V$ can be put to diagonal form. Define the variables:

\begin{align}
2\tau_1^2&:=V_{11}+1 \qquad 2\tau_2^2:=V_{22}+1\\
A&=\frac{1}{2\tau_1^2}+\frac{1}{2\tau_1^2} \qquad B=\frac{x}{\sqrt{2}\tau_2^2}+\frac{i p}{\sqrt{2}\tau_2^2}\\
C&=-\frac{1}{4\tau_2^2}+\frac{1}{4\tau_2^2} \qquad D=-\frac{x^2}{2\tau_2^2}-\frac{p^2}{2\tau_2^2}
\end{align}

Then the probability of photon number $n$ is

\begin{align}\label{thermphot}
p(n,\mathbf m, &V):=\tr{\hat \rho \ketbra{n}{n}}\nonumber\\
&=\frac{e^{D}}{\tau_1\tau_2}\sum_{i=0}^{n}\frac{n! (1-A)^i |C|^{n-i}}{i![(n-i)!]^2}\left|H_{m}\left(\frac{i B}{2\sqrt C}\right)\right|^2.
\end{align}

\end{document}